\documentclass[twocolumn,pre,aps,a4paper,floatfix]{revtex4-1}

\pdfoutput=1

\usepackage{graphicx}
\usepackage{bm}
\usepackage{color}
\usepackage{amsmath}
\usepackage{amssymb}
\usepackage{dsfont}
\usepackage{color}
\usepackage{calc}
\usepackage{hyperref}

\newcommand{\lla}{\left\langle}
\newcommand{\rra}{\right\rangle}

\definecolor{red}{rgb}{1,0,0}

\graphicspath{{./figures/}}

\begin{document}


\title{Thermostat for non-equilibrium  multiparticle collision dynamics simulations}

\author{Chien-Cheng Huang}
\author{Anoop Varghese}
\author{Gerhard Gompper}
\author{Roland G. Winkler}
\affiliation{%
Institute of Complex Systems and Institute for Advanced Simulation, Forschungszentrum J\"{u}lich, D-52425 J\"{u}lich, Germany}%

\date{\today}

\begin{abstract}
Multiparticle collision dynamics (MPC), a particle-based mesoscale simulation technique for complex fluid, is widely employed in non-equilibrium simulations of soft matter systems.  To maintain a defined thermodynamic state, thermalization of the fluid is often required for certain MPC variants. We investigate the influence of three thermostats on the non-equilibrium properties of a MPC fluid under shear or in Poiseuille flow. In all cases, the local velocities are scaled by a factor, which is either determined via a local simple scaling approach (LSS),  a Monte Carlo-like procedure (MCS), or by the Maxwell-Boltzmann distribution of kinetic energy (MBS).  We find that the various scaling schemes leave the flow profile unchanged and maintain the local temperature well. The fluid viscosities extracted from the various simulations are in close agreement.  Moreover, the numerically determined viscosities are in remarkably good agreement with the respective theoretically predicted values. At equilibrium, the calculation of the dynamic structure factor reveals that the MBS method closely resembles an isothermal ensemble, whereas the MCS procedure exhibits signatures of an adiabatic system at larger collision-time steps.
Since the velocity distribution of the LSS approach is non-Gaussian, we recommend to apply the MBS thermostat, which has been shown to produce the correct velocity distribution even under non-equilibrium conditions.

\end{abstract}
\maketitle

\section{Introduction}

Multiparticle collision dynamics (MPC) is a particle-based mesoscale simulation technique for fluids which has been introduced about a decade ago \cite{male:99,male:00,kapr:08,gomp:09}. By now, it has been developed into one of the major simulation techniques for complex fluids and has been applied to a broad range of soft matter systems. Examples cover equilibrium colloid
\cite{gomp:09,male:99,kapr:08,lee:04,hech:05,padd:06,goet:07,pete:10,whit:10,fran:11,belu:11} and polymer~\cite{gomp:09,kapr:08,male:00.1,ripo:04,muss:05,huan:10,huan:13} solutions and, more importantly,  non-equilibrium systems such as  colloids \cite{lamu:01,alla:02,wink:04,padd:04,ripo:08,wyso:09,goet:10.1,sing:11},  polymers
\cite{webs:05,ryde:06,ripo:06,cann:08,fran:08,niko:10,huan:10,fedo:12,huan:12.1,chel:12,jian:13},
vesicles \cite{nogu:04}, and cells \cite{nogu:05,mcwh:09} in flow
fields, colloids in viscoelastic fluids \cite{tao:08,ji:11,kowa:13}, as well as of self-propelled spheres \cite{ruec:07,goet:10,ming:11}, rods \cite{kapr:08,elge:09}, and other microswimmers \cite{earl:07,elge:10,reig:12,thee:13,elge:15}. Moreover, extensions have been proposed to fluids with non-ideal equations of state \cite{ihle:06} and mixtures \cite{tuez:07}.

The MPC algorithm consists of two discrete steps---streaming and collision---, and shares many features with the Direct Simulation Monte Carlo (DSMC) approach \cite{bird:94}. Although space is discretized into cells to define the multiparticle collision environment, both the spatial coordinates and the velocities of the particles are continuous variables. Hence, the algorithm exhibits unconditional numerical stability and satisfies a H-theorem \cite{male:99,ihle:03,gomp:09}.

MPC refers to a class of algorithms, which differ in their collision rules \cite{gomp:09}. In the original version of MPC, denoted as stochastic rotation dynamics (SRD) \cite{male:99,ihle:01}, collisions consist of a stochastic rotation of the relative velocities of the particles in a collision cell. Other algorithms assign Maxwellian distributed random relative velocities to the particles in a collision cell at every collision step, such that the momentum of the collision cell  is conserved (MPC-AT) \cite{alla:02,goet:07,nogu:07}.
MPC defines a discrete-time dynamics which has been shown to yield the correct long-time hydrodynamic behavior \cite{kapr:08,gomp:09,huan:12}. A consequence of the discrete dynamics is that the transport coefficients depend explicitly on the collision-time interval \cite{male:00,kapr:08,gomp:09,ihle:05,ihle:01,kiku:03,nogu:08,wink:09}, which in turn
permits control of fluid properties.

In many non-equilibrium systems, temperature has to be controlled to ensure a stationary state. A defined temperature is inherent in the MPC-AT approach \cite{alla:02,goet:07,nogu:07}, but an additional mechanism has to be provided for the MPC-SRD version, since it conserves energy. Various constant temperature simulation schemes have been proposed \cite{alle:87,fren:02,ande:80,heye:83,hail:83,bere:84,evan:84,nose:84,hoov:85,bulg:90,wink:92.2,wink:95,buss:07,huan:10.1}; not all of them ensure that a canonical ensemble is achieved and not all of them conserve momentum.

Under equilibrium conditions,
momentum can be conserved by velocity scaling schemes \cite{heye:83,hech:05,buss:07,gomp:09,huan:10.1}.
In its simplest form, velocity scaling keeps the kinetic energy of a system at the desired value by multiplying the velocities of all particles by the same factor \cite{huan:10.1}. This corresponds to an isokinetic rather that an isothermal ensemble. As far as MPC is concerned, a local cell-level scaling scheme (LSS) can be implemented, where a scale factor is determined for every cell independently. To achieve a canonical distribution of kinetic energies, more sophisticated cell-level approaches have been proposed based on a Monte Carlo criterion \cite{heye:83,schw:95,hech:05}, which we denote as Monte Carlo scaling (MCS), or by exploiting the appropriate distribution of the kinetic energy (Gamma distribution),   denoted as Maxwell-Boltzmann scaling (MBS)  \cite{huan:10.1}.

As is well known, under non-equilibrium conditions an inappropriate thermostat may introduce a bias into systems with an {\em a priori} unknown velocity profile \cite{evan:86}. To prevent a bias in MPC simulations, the relative particle velocities with respect to the center-of-mass velocity of a collision cell are scaled, which yields a profile unbiased thermostat (PUT) \cite{nogu:08,huan:10.1} and  renders a global scaling scheme inappropriate. However, recent detailed MPC simulation studies with a particular collision rule  indicate a substantial interference of certain thermostats with the flow field \cite{boli:12}. The comparison of viscosities, extracted from the parabolic flow profiles of Poiseuille flows, yields surprisingly large deviations between the values extracted from simulations and those determined theoretically. To avoid such an effect, Ref.~\cite{boli:12} suggests to exclude the flow and shear directions from thermostatting. Since, the large deviations are rather unexpected, we perform non-equilibrium MPC-SRD simulations with the ``standard'' three-dimensional collision rule in order to unravel the underlying cause.

A suitable thermostat is essential for accurate and reliable simulation results, and thermostats failing for flow fields like shear or Poiseuille flow are obviously inadequate for more complex flows. Hence, studies of the reliability of a thermostat combined with a particular MPC collision rule are important.  In this article, we characterize the influence of the  LSS,  MCS, and MBS thermostats on the velocity profile of simple shear and Poiseuille flow.  Thereby, we determine the viscosity for various collision times and compare it with the theoretical  expression.  As a reference, we also determine the fluid viscosity  at equilibrium via the transverse fluid-velocity autocorrelation function in Fourier space without any thermostat \cite{huan:12}. We find very good agreement between the velocity profiles for the various thermostats. The differences between the relative viscosities are below $2\%$  and thus agree with each other within the accuracy of the simulations. The analytically derived expression of the viscosity of a MPC fluid is based on some approximations, specifically the molecular-chaos assumption. Hence, it is not {\em a priory}  evident to which extent it describes simulation results. We find a remarkably good agreement between the viscosity extracted from simulations and the theoretical expression, which emphasizes the importance of the theoretical result. Moreover, we calculate the equilibrium dynamic structure factor of fluids thermalized by the various thermostats. For systems with temperature controlled by LSS or MBS, the structure factor lacks a Rayleigh line, which corresponds to an isothermal ensemble \cite{boon:80,hans:86}. For systems with a MCS thermostat, a Rayleigh peak appears for larger collision-time steps, hence, no isothermal system is simulated.

The article is organized as follows. In Sec.~\ref{sec:model} the model and simulation approach are introduced. Section ~\ref{sec:thermostats} describes the various thermostats. Results for equilibrium
hydrodynamic fluctuations, specifically transverse velocity autocorrelation functions and dynamics structure factors, are presented in Sec.~\ref{sec:hydro_fluct}. The results of non-equilibrium simulations are presented in Sec.~\ref{sec:results} and various implications are discussed in Sec.~\ref{sec:discussion}. Finally, Sec.~\ref{sec:conclusion} summarizes our findings.

\section{Model and Methods}  \label{sec:model}

\subsection{Multiparticle Collision Dynamics} \label{sec:mpc_method}

We consider a  MPC fluid of $N$ point particles of mass $m$. Without external field, the particles  move ballistically during the streaming step, i.e., their positions  ${\bm r}_i$ are updated according to
\begin{align} \label{streaming}
{\bm r}_i (t+h) = {\bm r}_i(t) + h {\bm v}_i(t) ,
\end{align}
where ${\bm v}_i(t)$ is the velocity of particle $i$ at time $t$ and $h$ is the collision-time step.
In the collision step,  the simulation box is partitioned into cubic collision cells of side length $a$. In the SRD version of MPC,  the relative velocity of each particle, with respect to the center-of-mass velocity of the particular cell, is rotated by a fixed angle $\alpha$ around a randomly oriented axis. Hence, the velocity after a MPC step is
\begin{align}
{\bm v}_i(t+h)= {\bm v}_i(t)+\left[\mathbf{R}(\alpha)-\mathbf{E} \right] ({\bm v}_i(t)-{\bm v}_{cm}(t)),
\end{align}
where $\mathbf{R}(\alpha)$ is the rotation matrix, $\mathbf{E}$ is the unit matrix, and
\begin{equation}
{\bm v}_{cm} = \frac{1}{N_c} \sum^{N_c}_{j=1}{\bm v}_j ,
\end{equation}
is the center-of-mass velocity of the $N_c$ particles contained in the cell of particle $i$ \cite{male:99,male:00,kapr:08,gomp:09}. The random orientation of the rotation axis is chosen independently at every collision step and for every cell \cite{huan:10.1}.
To insure Galilean invariance, a random shift is performed at every collision step \cite{ihle:01}.
In a collision step, mass, momentum, and energy are conserved which leads to the build-up of correlations in the particle motion and gives rise to hydrodynamic interactions.

\subsection{Boundary conditions} \label{sec:boundary_conditions}

We typically apply three-dimensional periodic boundary conditions with a cubic simulation box of side length $L$ and volume $V=L^3$. In many systems, e.g., in simulations of Poiseuille flow, solid walls are present, commonly with no-slip boundary conditions. The discretization into collision cells requires particular measures to ensure the no-slip condition. We follow the approach suggested in Ref.~\cite{lamu:01}, which applies the bounce-back rule, a random shift of the collision lattice in all spatial directions, and partial filling of surface cells by phantom particles.
For systems with parallel walls, the random shift perpendicular to the walls is implemented as follows \cite{wink:09}. Without random shift, collision-cell boarders are chosen to coincide with the respective wall. To enable a random shift, an additional layer of empty collision cells is added in one of the  walls. In a random shift, the whole collision lattice is then shifted by a uniformly distributed random displacement $\in [0,a]$. The  typically appearing partially occupied cells at the walls  cause a violation of the no-slip boundary condition, since the average fluid velocity parallel to the surfaces is  no-longer zero in the surface cells \cite{lamu:01}. To restore no-slip boundary conditions, usually a phantom  particle is added to every cell intersected by a wall and occupied by  $N_{sc}$ fluid particles  smaller than the average number of particles $\lla N_c \rra$, such that the average particle density is restored. However, this does not completely prevent slip, because the average center-of-mass position of all particles in a collision cell---including the phantom particle, which is placed in the center of the wall-occupied part of the collision cell---does not coincide with the wall. In order to fully account for the no-slip boundary condition, we adopt the following modification of the original approach \cite{wink:09}. To treat a wall cell on the same basis as a cell in the bulk, i.e., the number of fluid particles satisfies a Poisson distribution with the average $\lla  N_c \rra$, we take fluctuations in the particle number into account by adding $N_{sp}$ particles to every cell partially occupied by a wall such that $ \lla N_{sp} + N_{sc} \rra = \lla N_c \rra$. The momentum ${\bm P}$ of all phantom particles of a cell is taken from the Maxwell-Boltzmann distribution with the variance $ mN_{sp}k_BT$ and, at equilibrium, zero average. There are various ways to determine the number $N_{sp}$. For a system with two parallel walls, we suggest to use the number of fluid particles in the surface cell intersected by the opposite wall. The average of the two numbers is equal to $\lla N_c\rra$. Alternatively, $\lla N_{sp} \rra$ can be taken from a Poisson distribution with average $\lla N_c \rra$ accounting for the fact that there are already $N_{sc}$ particles in the cell. Collisions are then performed with all the particles in the cells.
The center-of-mass velocity of the particles in a boundary cell is
\begin{align}
{\bm v}_{cm} = \frac{1}{N_{sc}+ N_{sp}} \left( \sum_{i=1}^{N_{sc}} m {\bm v}_i + {\bm P} \right) .
\end{align}
Instead of a single phantom-particle momentum and a single extra collision-lattice layer, an additional collision-cell layer can be added in every wall and explicitly be filled with randomly placed phantom  particles with Maxwellian distributed velocities in every collision step. If the layers are sufficiently large, the fluctuations of the particle number in a collision cell are close to that of a bulk cell.

Other approaches for no-slip boundary conditions have been suggested and analyzed \cite{inou:02,padd:06,hech:05,whit:10,boli:12}. Some of them do not include phantom particles. However, based on our experience, we consider the approach with phantom particles as most appropriate for no-slip boundary conditions, which yields the correct hydrodynamic behavior not only for  solid walls, but also for solid particles immersed in a MPC fluid.

The presence of external fields may require an adaptation and modification of the boundary conditions to ensure the no-slip requirement. For simple shear such a adjustment is described in Ref.~\cite{wink:09}, and for Poiseuille flow in Ref.~\cite{boli:12}. We will come back to this aspect in Sec.~\ref{sec:poiseuille_flow}.

\subsection{Shear Flow} \label{sec:shear_flow}

Shear flow is implemented by Lees-Edwards periodic boundary conditions \cite{lees:72,alle:87}, which yields a linear velocity profile \cite{wink:09}. The shear viscosity $\eta$ of the fluid follows from  the stress tensor $\sigma_{xy}$ via $\eta= \sigma_{xy} /\dot \gamma$ for shear along the $x$-axis and the gradient direction along the $y$-axis of the Cartesian reference frame; $\dot \gamma$ denotes the shear rate.  As shown in Ref.~\cite{wink:09},
the instantaneous stress of the MPC fluid is given by
\begin{align}   \label{stress_tensor}
\sigma_{xy}^i  & = - \frac{1}{V} \sum_{i=1}^{N}  m  \hat v_{i x} \hat v_{i y}
 - \frac{\dot \gamma h}{2V} \sum_{i=1}^{N} m v_{iy}^2
 -\frac{1}{Vh}  \sum_{i=1}^{N} \Delta p_{i x} r_{iy} ,
\end{align}
hence, $\sigma_{xy} = \lla \sigma_{xy}^i \rra$. $\hat v_{i \alpha}$, $\alpha \in \{x,y,z\}$, is the velocity after streaming, but before  collision, whereas $v_{i \alpha}$ is the velocity after collision, but before streaming, and $\Delta p_{i\alpha}$ is the momentum change of particle $i$ during a collision. Here, ${\bm r}_i$ is the position of the particle in the grid-shifted frame.  Note that due to Lees-Edwards boundary conditions, all particles are inside of the primary periodic box and the velocities along the flow direction are correspondingly adjusted \cite{wink:09}.

\subsection{Poiseuille  Flow} \label{sec:poiseuille_flow}

 For the Poiseuille flow simulations, we confine the fluid between two solid walls, which are parallel to the $xz$-plane of the Cartesian reference frame,  and apply periodic boundary conditions along the $x$- and $z$-axis. Flow is induced by  a constant force $mg$ acting on every fluid particle. Therefore, the particle velocities and positions are updated according to
 \begin{align}
 \hat v_{ix} (t+h) & = v_{ix} + g h ,\\
 \hat r_{ix} (t+h) & = r_{ix}(t) + v_{ix}(t) h + \frac{1}{2}g h^2
 \end{align}
 along the flow direction. Note that the circumflex indicates quantities after streaming but before collision.
 To satisfy the no-slip boundary condition, we apply the bounce-back rule during streaming and take into account phantom particles during collisions. We consider two variants for the calculation of a phantom-particle momentum. In the simpler version, the average momentum $\lla{\bm P}\rra$ is set to zero. However, this implies a residual slip. Following the  proposition for shear flow in the presence of walls in Ref.~\cite{wink:09}, we  assign a finite mean velocity to every phantom particle according to its position in the collision cell relative to the wall-fluid interface. Thereby, we place a phantom particle in the center of the part of the collision cell inside a wall. The average velocity itself is determined by the desired parabolic flow profile. A similar approach has been adopted in Ref.~\cite{boli:12}.

\subsection{Viscosity} \label{sec:viscosity}

Analytical expressions for the MPC fluid viscosity have been derived applying various methods \cite{male:00,kapr:08,gomp:09,ihle:05,ihle:01,kiku:03,pool:05,nogu:08,wink:09}. In general, the viscosity $\eta = \eta^k + \eta^c$ comprises a kinetic contribution $\eta^k$  due to streaming of the fluid particles, and a collisional contribution $\eta^c$.  For the latter, an exact expression can be derived, which is given by
\begin{align} \label{visc_coll}
\eta^{c} = \frac{N m a^2}{18 V h}  \left ( 1 -  \cos(\alpha) \right)
\left(1 - \frac{1}{\lla N_c \rra} \right)
\end{align}
for $\lla N_c \rra \gg 1$ in three dimensions. Here and in the following, we neglect fluctuations in the particle number in a collision cell, which is justified for average particle numbers $\lla N_c \rra >5$, since we omit a term of order $e^{-\lla N_c\rra}$. Due to correlations in the particle velocities, the kinetic contribution can only be derived within certain approximations. Applying the molecular chaos assumption, i.e.,  velocity correlations between different particles are neglected, the kinetic contribution is
\begin{align} \label{visc_kin}
\eta^{k} = \frac{N k_B T h}{2V}  \left[ \frac{5 \lla N_c \rra}{(\lla N_c \rra-1)(2 - \cos(\alpha) -  \cos (2\alpha)  )} - 1 \right] .
\end{align}
Evidently, the collisional contribution dominates for small  and the kinetic one for large collision-time steps, which corresponds to a fluid-like behavior in the first case and gas-like behavior in the second case as expressed by the Schmidt number \cite{ripo:05}.

In Ref.~\cite{ripo:05} and especially in Ref.~\cite{tuez:06} for two-dimensional systems, improved analytical expressions are provided for $\eta^k$ taking into account velocity correlations. It is important to note that fluid correlations yield a somewhat larger $\eta^k$ value than that predicted by  the molecular-chaos assumption.

The total viscosity $\eta = \eta^k + \eta^c$ is evidently dominated by $\eta^k$ for $h \to \infty$ and $\eta^c$ for $h \to 0$. Since $\eta^c$ is calculated without any approximation, $\eta$ provides an exact description for $h \to 0$. Moreover, the applied molecular chaos assumption applies well for $h \to \infty$. Hence, $\eta$ is well described quantitatively by the theoretical expression in both limits.

\subsection{Parameters}

All simulation are performed with the rotation angle $\alpha=130^{\circ}$, and the mean number of particles per collision cell $\lla N_c \rra  = 10$. Length and time are measured in units of the collision cell size $a$ and   $\tau=\sqrt{ma^2/(k_BT)}$, respectively,  where $T$ is the temperature and $k_B$ the Boltzmann constant. Various collision times between $h = 0.1 \tau$ and $h=3 \tau$ are considered to cover the collisional-dominated as well as the streaming-dominated regime.  The size of the cubic simulation box is set to $L=30a$ if not otherwise stated.  For the shear flow simulation the choose the shear rate $\dot \gamma \tau = 10^{-2}$, and for the Poiseuille flow simulations, we set $g=10^{-3} a/\tau^2$.

For an efficient simulation of the MPC fluid dynamics, we exploit a graphics processor unit (GPU) based version of the simulation code \cite{west:14}.

\section{Thermostat} \label{sec:thermostats}

We perform simulations applying the thermostats mentioned in the introduction, namely local simple scaling (LSS) \cite{huan:10.1}, Monte Carlo scaling (MCS) as suggested in Ref.~\cite{hech:05}, and Maxwell-Boltzmann scaling (MBS) \cite{huan:10.1}. In all case, the relative velocities
$\Delta {\bm v}_i = {\bm v}_i - {\bm v}_{cm}$ of the particles in a collision cell are scaled by a constant factor $\xi$,  which typically differs for every cell and collision-time step. Hence, the relative velocities after collision $\Delta {\bm v}_i' $ are given by $\Delta {\bm v}_i' = \xi \Delta {\bm v}_i$.  Since the total relative momentum of a collision cell is zero, such a scaling leaves the total momentum of a cell unchanged.\\

{\em Simple Scaling---}In the simple scaling approach, the scale factor $\xi$ is chosen as
\begin{align}
\xi = \sqrt{\frac{ 3 (N_c-1) k_B T }{2 E_k} } ,
\end{align}
with the  kinetic energy of a collision cell
\begin{align} \label{kin_energy}
E_k = \frac{1}{2} \sum_{i=1}^{N_c} m \Delta {\bm v}_i^2 .
\end{align}
The factor $N_c-1$ accounts for the fact that $E_k$ is calculated in the center-of-mass reference frame of a collision cell.  Strictly speaking, LSS conserves the average kinetic energy rather than the temperature \cite{huan:10.1}. This implies that the energy fluctuations are incompatible with that of an isothermal ensemble and the distribution of velocities in a collision cell is not Maxwellian.\\

{\em Monte Carlo Scaling---}We implement the Monte Carlo scaling method in the version proposed in Ref.~\cite{hech:05}, which satisfies detailed balance in contrast to earlier versions \cite{heye:83}. In brief, a factor $\epsilon$ is randomly chosen in the interval $[0,\zeta]$ and $\xi$ is set to either $1+\epsilon$ or $1/(1+\epsilon)$, each with the probability $1/2$. The velocity scaling itself is performed following a Monte Carlo-type criterion, with the probability $p_A= \mathrm{min}[1,A]$, where \cite{hech:05,boli:12}
\begin{align}
A= \xi^{3(N_c-1)}\exp\left( -(\xi^2-1) E_k/ k_BT\right) .
\end{align}
The choice of $\zeta \in [0.05$,$0.3]$ and the frequency of scaling determine the relaxation time to approach the desired temperature $T$. We set $\zeta = 0.1$.
The method has been shown to yield the correct velocity distribution \cite{hech:05} and has successfully been applied in various simulation studies \cite{hech:05,boli:12,hech:06,hech:07,hech:07.1}. \\

{\em Maxwell-Boltzmann Scaling---}In the Maxwell-Boltz\-mann scaling approach, the known distribution of the kinetic energy of the MPC ideal-gas particles is exploited to determine the scale factor \cite{huan:10.1}.  Thereby,  the distribution of the kinetic energy is given by ($\Gamma$ distribution)
\begin{align}  \label{distr_kin_energy}
P(E_k) = \frac{1}{E_k \Gamma (f/2)}\left(\frac{E_k}{k_BT}\right)^{f/2}
\exp \left(- \frac{E_k}{k_B T} \right) .
\end{align}
Here, $f = 3(N_c-1)$ is the number of degrees of freedom of the fluid particles in the considered collision cell, and $\Gamma(x)$ is the gamma function. In the limit $f \to \infty$, the $\Gamma$ distribution turns into a Gaussian function with mean $\left\langle E_k \right\rangle = f k_BT/2$ and variance $f (k_BT)^2 /2$. At every collision, a random kinetic energy $\hat E_k$ is taken from the distribution function (\ref{distr_kin_energy}) for every collision cell and the respective scale factor for the velocities is set to
\begin{align} \label{scal_factor_mbs}
\xi = \sqrt{\hat E_k/E_k} ,
\end{align}
with the kinetic energy $E_k$ of Eq.~(\ref{kin_energy}). Taking the average of the kinetic energy after scaling, we obtain
\begin{align}
\lla E_{k} \rra = \frac{1}{2} \sum_{i=1}^{N_c} m \lla \Delta {\bm v}_i'^2\rra = \lla \frac{\xi^2}{2} \sum_{i=1}^{N_c} m \Delta {\bm v}_i^2 \rra = \lla \hat E_k \rra .
\end{align}
Hence, the average of the kinetic energy of a collision cell is equal to the desired mean of the distribution function (\ref{distr_kin_energy}). Note that in Ref.~\cite{boli:12}, a different scale factor is provided, which may simply be a misprint, otherwise the factor would not provide the correct average kinetic energy.

As has been shown in Ref.~\cite{huan:10.1}, the MBS approach yields the correct distribution function of the particle velocities at the collision cell level, even for strong external fields, whereas LSS fails even at equilibrium.

When phantom particles are taken into account, the MPC particles next to a wall are thermalized by the phantom particles. For sufficiently weak (external) fields, this energy exchange suffices to control the temperature in the whole system. For strong fields, the energy transport is not fast enough to ensure the desired temperature over the whole system \cite{huan:10.1}. Here, one of the additional thermostating schemes has to be applied.

\section{Results: Hydrodynamic Fluctuations} \label{sec:hydro_fluct}

The hydrodynamic properties of a MPC fluid coincide with those of the linearized fluctuating hydrodynamic equations for sufficiently large length and time scales \cite{male:99,kapr:08,pool:05,ihle:05,ihle:06,huan:12}. Hence, we can use hydrodynamic correlation functions, on the one hand, to extract the fluid transport coefficients from equilibrium velocity autocorrelation functions, and, on the other hand, to verify the kind of simulated ensemble.

Within the linearized Navier-Stokes equations \cite{boon:80,hans:86}, the transverse hydrodynamic (shear) modes are independent of the longitudinal (acoustic, entropy) modes \cite{boon:80}. For an adiabatic system, i.e., a system in which the energy of the fluid is locally conserved, the longitudinal modes are coupled. In particular, the temperature (or entropy) fluctuations are coupled to the density and longitudinal velocity fluctuations \cite{boon:80,hans:86}. In contrast, in an isothermal system, temperature fluctuations are suppressed and controlled by the (local) thermostat, which implies a decoupling of the density and velocity fluctuations from the equation of the temperature fluctuations. The respective modifications of the transport properties are reflected in the density autocorrelation function, e.g., the dynamic structure factor $S({\bm k},\omega)$.

In the following two subsections, we will address the transverse velocity correlation function and the density fluctuations via the dynamic structure factor.

\subsection{Transverse Velocity Correlation Function}

\begin{figure}[t]
\begin{center}
\includegraphics[width=\columnwidth,angle=0]{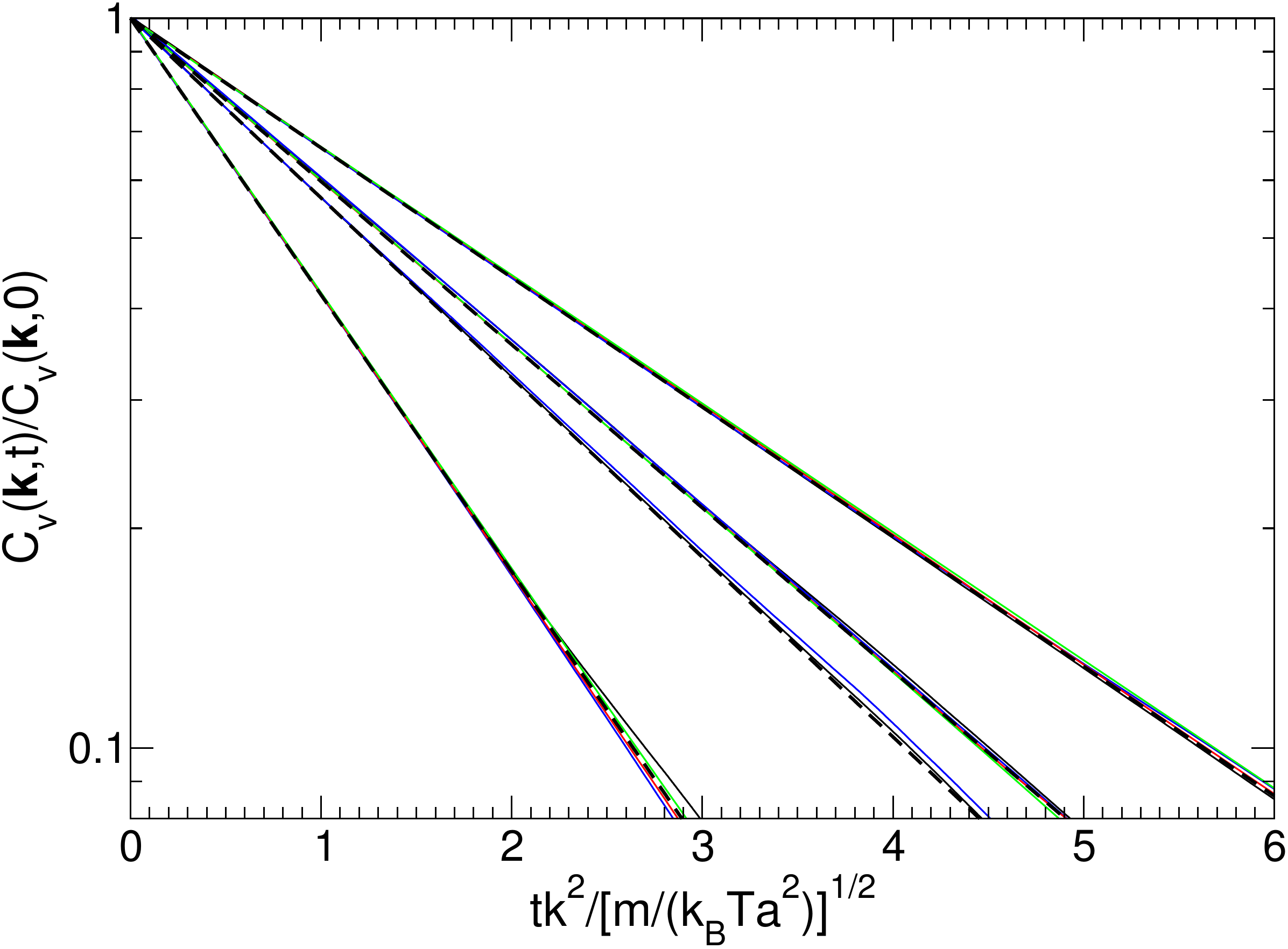}
\caption{(Color online) Transverse velocity autocorrelation functions of a MPC fluid for the  collision-time steps $h/\tau=0.1$, $1.0$, $0.2$, and $0.5$ (left to right). The $k$-values are $k=2\pi n/L$, with $L=30a$ and $n=1$, $2$, $3$, and $4$.  The fit of the exponential function (\ref{corr_trans}) (dashed lines) yields  the kinematic viscosities presented in Table~\ref{table1}.}
\label{fig:corr_trans}
\end{center}
\end{figure}

The MPC fluid viscosity can be determined by equilibrium simulations,  independent of any thermostat,  from the transverse velocity correlation function in Fourier space \cite{ihle:03,ihle:03.1,huan:12}. For the considered periodic system, the velocity in $k$-space is defined as \cite{huan:12,duen:93.1}
\begin{align}
 {\bm v}({\bm k,t})= \sum_{i=1}^N {\bm v}_i(t) e^{i{\bm k} \cdot {\bm r}_i(t)} ,
\end{align}
with $k_{\beta}= 2 \pi n_{\beta}/L$, $\beta \in \{x,y,z\}$, $n_{\beta}\in \mathbb{Z}$, and ${\bm k} \neq 0$.
From the linearized hydrodynamic equations \cite{land:59,hans:86,boon:80,huan:12,thee:14},  the normalized transverse velocity correlation function $C_v({\bm k},t)$ is obtained, which decays exponential as
\begin{align} \label{corr_trans}
C_v({\bm k},t) = \frac{\lla {\bm v}^T({\bm k},t) \cdot  {\bm v}^T({-\bm k,0)}\rra}{\lla {\bm v}^T({\bm k},0)^2\rra }  = e^{- {\bm \nu k}^2 t} ,
\end{align}
where $\nu=\eta/\varrho$ is the kinematic viscosity and $\varrho$ the mass density.

Figure~\ref{fig:corr_trans} shows examples of autocorrelation functions for various collision-time steps extracted from simulations without a thermostat. Evidently, the obtained $C_v(t)$ decay exponentially. A fit with the exponential function  (\ref{corr_trans})  yields the viscosities
listed in Table~\ref{table1}. These values are slightly larger than the theoretical values calculated according to Eqs.~(\ref{visc_coll}) and (\ref{visc_kin}), which is a consequence of the applied approximations in the derivation of the theoretical expressions.

We performed various simulations applying a thermostat and calculated $C_v({\bm k},t)$. Within the accuracy of the results, we did not detect any difference between simulations with and without thermostat.

For the sake of completeness, we like to mention that Fourier transformation of  the correlation function $C_v(\bm k,t)$ in Eq.~(\ref{corr_trans}) yields the well-known long-time tail, characteristic for hydrodynamic correlations \cite{alde:70,zwan:70,huan:12}. Further details are presented in Ref.~\cite{huan:12}.

\subsection{Dynamic Structure Factor}

The dynamic structure factor is defined as
\begin{align} \label{eq:dynamic_sf}
S({\bm k},\omega) = \frac{1}{N} \int_{-\infty}^{\infty} \lla \delta \rho({\bm k},t) \delta \rho(-{\bm k},0) \rra e^{-i\omega t} dt
\end{align}
in terms of the (number) density fluctuations $\delta \rho({\bm r},t) = \rho({\bm r},t) -\rho$ \cite{boon:80,hans:86,tuez:06,hija:11,huan:10}, where $\rho$ denotes the mean density and
\begin{align}
\rho ({\bm k},t) = \sum_{i=1}^N e^{i{\bm k} \cdot {\bm r}_i(t)} .
\end{align}
Explicitly, the normalized dynamic structure factor $\tilde S$ ($\int \tilde S d\omega =1$) of an adiabatic system for small $k$ values is given by \cite{boon:80}
\begin{align} \label{eq:dynamic__sf_adiabatic}
2 \pi & \tilde S({\bm k},\omega)  = \frac{\gamma -1}{\gamma} \frac{2 D_Tk^2}{\omega^2 + (D_T k^2)^2}
\\ \nonumber & + \frac{1}{\gamma} \left[ \frac{\Gamma_s k^2}{(\omega + c_sk)^2 + (\Gamma_s k^2)^2} + \frac{\Gamma_s k^2}{(\omega -c_sk)^2 + (\Gamma_s k^2)^2}   \right] \\ \nonumber &
+ \frac{1}{\gamma}\left[ \Gamma_s + (\gamma -1)D_T \right] \frac{k}{c_s} \\ \nonumber
& \times \left[ \frac{\omega + c_s k}{(\omega + c_sk)^2 + (\Gamma_s k^2)^2}  - \frac{\omega - c_sk}{(\omega -c_sk)^2 + (\Gamma_s k^2)^2}   \right] ,
\end{align}
where $c_s=\sqrt{\gamma k_B T/m}$ is the adiabatic velocity of sound, $D_T$ the thermal diffusion coefficient, $\Gamma_s$  the sound attenuation factor, and $\gamma$ the adiabatic index. More definitions and the relation with the MPC parameters are provided in Appendix \ref{appendix}.
The expression for an isothermal system follows by setting  $D_T=0$ and $\gamma=1$ \cite{hija:11}:
\begin{align} \label{eq:dynamic__sf_isothermal}
2 \pi \tilde S({\bm k},\omega)  = \frac{\Gamma k}{c} & \left[ \frac{2 c k + \omega }{(\omega + ck)^2 + (\Gamma k^2)^2} \right. \\ \nonumber  & + \left.  \frac{2 ck - \omega}{(\omega - ck)^2 + (\Gamma k^2)^2} \right].
\end{align}
Here, $c=\sqrt{k_B T/m}$ denotes the isothermal speed of sound and $\Gamma$ the isothermal sound attenuation factor (see Appendix \ref{appendix}). Note that the structure factor is related to the longitudinal velocity autocorrelation function  via \cite{boon:80,hans:86}
\begin{align}
\frac{1}{N} \int \langle v^L({\bm k}, t) v^L(-{\bm k}, 0)  \rangle e^{-i \omega t} dt = \left(\frac{\omega}{\varrho |{\bm k}|}\right)^2 S({\bm k},\omega) .
\end{align}
This correlation function lacks a Rayleigh line due to the appearing frequency ($\omega^2$) on the right-hand side.

\begin{figure}[t]
\begin{center}
\includegraphics[width=\columnwidth,angle=0]{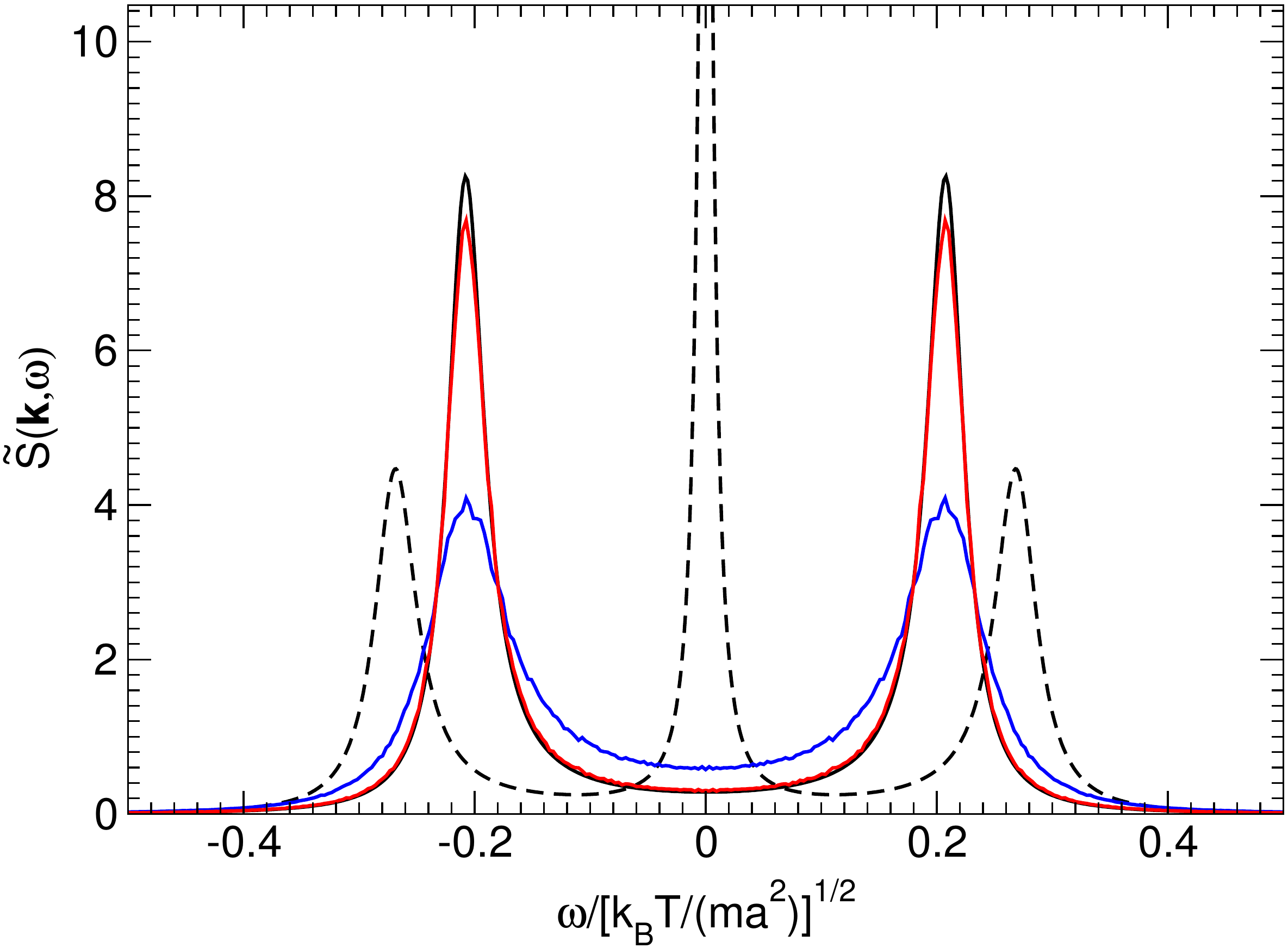}
\caption{(Color online) Normalized dynamic structure factors for $h=0.1\tau$. The solid line with the smallest peaks (blue) correspond to the MCS thermostat, the line with the next larger peaks (red) to the MBS thermostat, and the curve with the most pronounced  peaks (black) is the theoretical structure factor of an isothermal system (Eq.~(\ref{eq:dynamic__sf_isothermal})). The dashed curve indicates the  theoretical structure factor of an adiabatic system (Eqs.
(\ref{eq:dynamic__sf_adiabatic})).
The system size is $L/a=30$. }
\label{fig:longitudinal_01}
\end{center}
\end{figure}

\begin{figure}[t]
\begin{center}
\includegraphics[width=\columnwidth,angle=0]{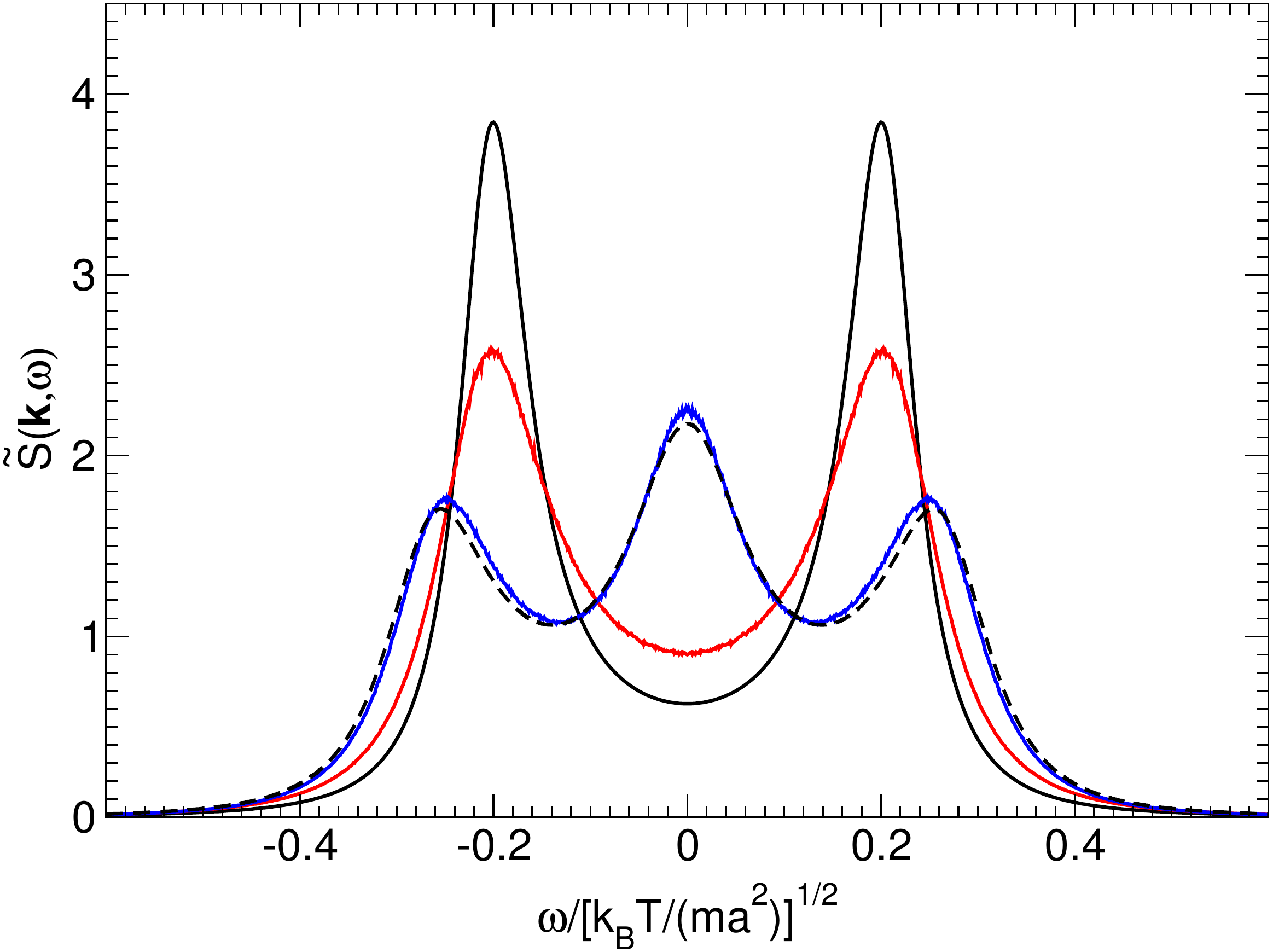}
\caption{(Color online) Normalized dynamic structure factors for $h=3.0\tau$. The solid line with the smallest peaks (blue) correspond to the MCS thermostat, the line with the next larger peaks (red) to the MBS thermostat, and the curve with the most pronounced  peaks (black) is the theoretical structure factor of an isothermal system (Eq.~(\ref{eq:dynamic__sf_isothermal})). The dashed curve indicates the  theoretical structure factor of an adiabatic system (Eqs.
(\ref{eq:dynamic__sf_adiabatic})
The system size is $L/a=30$.}.
\label{fig:longitudinal_30}
\end{center}
\end{figure}

Figures \ref{fig:longitudinal_01} and \ref{fig:longitudinal_30} provide examples of $\tilde S({\bm k}, \omega)$ for the collision times $h =0.1 \tau$ and $3.0 \tau$, respectively, and the MBS and MCS scaling schemes. For LSS, we obtain the identical structure factors as for MBS within the accuracy of the simulations. For the short collision-time step (Fig.~\ref{fig:longitudinal_01}), two Brillouin lines are present at the frequencies $\omega \approx \pm c k$. No central Rayleigh line is present, hence, there are no temperature fluctuations. The simulation result of the MBS thermostat is in very close agreement with the theoretical prediction, whereas the height of the Brillouin peaks is smaller for the MCS thermostat, but the peak positions correspond to those of an isothermal system.

Similarly, for the simulations with $h = 3.0 \tau$ (Fig.~\ref{fig:longitudinal_30}), the Brillouin lines of the MBS thermostat correspond to those of an isothermal system, although the peak height is somewhat smaller than that of an isothermal system. In contrast, the structure factor for the MCS thermostat is close to  the theoretical expression of an adibatic system. The Brillouin peaks shift to the frequencies $\omega \approx \pm \sqrt{\gamma} c k$, corresponding to adibatic sound propagation. More importantly, there is a central Rayleigh line.

Thus, the MCS thermostat at large collision-time steps is not reproducing an isothermal but rather an adiabatic system. This may not necessarily be a problem for temperature control, since the Monte Carlo procedure approaches the desired canonical velocity distribution in the limit of a large number of attempts; however, the temperature fluctuations are not correct locally. In addition, typically collision-time steps $h <0.2 \tau$ are used to simulate fluids. Here, a nearly isothermal system is achieved for MCS.

The deviation between the theoretical structure factor of an isothermal system and the simulation data for MCS at $h=0.1\tau$ and MBS at  $h=3.0\tau$, respectively, indicates that neither method controls temperature perfectly locally for these time steps. We attribute the deviation from the isothermal dynamic structure factor to streaming of the MPC particles. For the MBS thermostat and the collision-time step $h=0.1\tau$, there is very little energy transport during streaming, and thus, the system closely resembles an isothermal ensemble. However, for $h=3.0 \tau$, there is a considerable energy transfer to nearby collision cells in the streaming step, which implies non-isothermal fluctuations. In case of the MCS method, velocity scaling occurs with a certain probability only, which leads to large displacements of particles without real temperature control. This is particularly pronounced for $h=3.0 \tau$, where heat is transferred  over large distances during streaming and gives rise to  adiabatic rather than isothermal fluctuations. The crossover from isothermal to adiabatic density fluctuations has been addressed in Ref.~\cite{hija:11}.

We finally would like to emphasize that the dynamic structure factor for systems with the MCS thermostat depends
on the parameter $\zeta$. Simulations with the ``extreme'' values  $\zeta = 0.05$ and $0.3$ lead to
slight shifts of the Brillouin lines and variation in the peak heights. However, for $h=3 \tau$, there is always a pronounced Rayleigh peak.

\section{Results: Non-equilibrium Simulations} \label{sec:results}

\begin{table}[htbp]
\caption{\label{table1}
Kinematic viscosities $\nu = \eta/\varrho$ and their deviations $\Delta\nu=(\nu-\nu^{th})/\nu^{th}\times 100\%$
from theoretical values $\nu^{th}$ [$\nu^{th} = \nu^k+\nu^c$, Eqs.~(\ref{visc_coll}) and
(\ref{visc_kin})] obtained from the velocity autocorrelation function (\ref{corr_trans}) and shear-flow
simulations for the thermalization methods:
local simple scaling (LSS), Monte Carlo scaling (MCS) \cite{hech:05}, and Maxwell-Boltzmann scaling (MBS)
\cite{huan:10.1}. The simulation box size is set to $L=60 a$ in the calculation of the VACF for $h/\tau = 1$, $2$, and $3$.}
\begin{ruledtabular}
\begin{tabular}{l|ccccccc}
 & $h/\tau$ & 0.1 & 0.2 & 0.5 & 1.0 & 2.0 & 3.0 \\
\hline
Theory & $\nu^{th}/(a^2/\tau)$& 0.870 & 0.508 & 0.407 & 0.568 & 1.014 & 1.486 \\
\hline
& $\nu/(a^2/\tau)$& 0.873 &  0.515 & 0.409 & 0.569 & 1.006 & 1.484  \\
\raisebox{1.5ex}[-1.5ex]{VACF}  & $\Delta \nu/\%$& 0.4 & 1.3  & 0.5 & 0.2 & -0.8 & -0.1 \\ \hline
 & $\nu /(a^2/\tau)$ & 0.872 & 0.517 & 0.411 & 0.571  & 1.017 & 1.492 \\
\raisebox{1.5ex}[-1.5ex]{LSS} & $\Delta \nu/\%$ & 0.2 & 1.7 & 0.9 & 0.4 & 0.4 & 0.4 \\ \hline
& $ \nu/(a^2/\tau)$& 0.869 & 0.515 & 0.412 & 0.571 & 1.016 & 1.493\\
\raisebox{1.5ex}[-1.5ex]{MCS} & $\Delta \nu /\%$& -0.1 & 1.4  & 1.2 & 0.4 & 0.3 & 0.5\\ \hline
 & $\nu/(a^2/\tau)$& 0.871 & 0.517 & 0.414 & 0.573 & 1.019 & 1.494  \\
\raisebox{1.5ex}[-1.5ex]{MBS} & $\Delta \nu/\%$& 0.1 & 1.8 & 1.5 & 0.8 & 0.5 & 0.5 \\
\end{tabular}
\end{ruledtabular}
\end{table}

We determine the fluid viscosity via non-equilibrium simulations in order to demonstrate that the viscosity is independent of the thermostat and, moreover, that the thermostat is not interfering with the flow.

\subsection{Shear Flow}

We perform shear-flow simulations for various collision times and the LSS, MCS, and MBS thermostat.  In all cases, we obtain a linear velocity profile, in agreement with the theoretical expectation. From the stress tensor values, we calculate the viscosities listed in Table \ref{table1}. The viscosities attained by the various thermostats are in close agreement with each other and are in remarkable agreement with the theoretical prediction. As expected, the simulation values are typically slightly larger than the theoretically determined viscosities. However, they agree within about $2\%$. The largest deviation appears for $h=0.2 \tau$. This supports our expectation that the theoretically derived expression for the viscosity agrees well with simulation results  for larger and smaller collision-time steps.
The agreement between the viscosities extracted from simulations is even better; the relative error is below $1\%$.

\begin{figure}[t]
\begin{center}
\includegraphics[width=\columnwidth,angle=0]{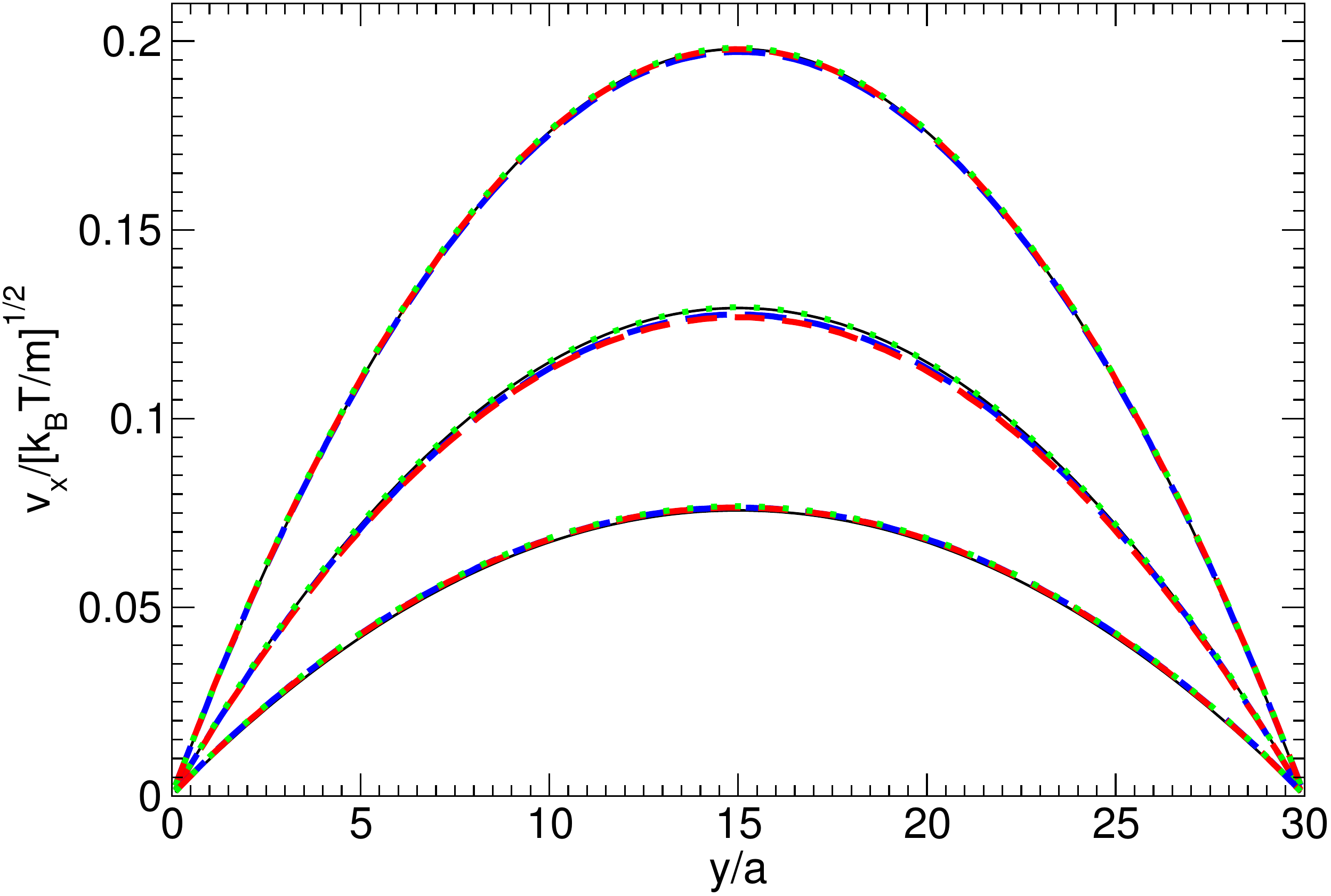}
\caption{(Color online) Poiseuille-flow velocity profiles of simulations with non-zero average phantom-particle momenta, i.e., no-slip boundary conditions, for the simple scaling (LSS) (short-dashed, red), the Monte Carlo scaling (MCS) (dotted, green), and the Maxwell-Boltzmann  (MBS) (dashed, blue) thermostat. The parabolic velocity profiles with the theoretically determined viscosities (cf. Table~\ref{table1}) and zero slip length are shown by solid lines (black). The collision-time steps are $h/\tau =1.0$, $0.1$, and $3.0$ (top to bottom).}
\label{fig:pois_flow}
\end{center}
\end{figure}

\begin{table}[tbp]
\caption{\label{table2} Kinematic viscosities $\nu$ and their deviations $\Delta \nu$ with respect to theoretical values extracted from Poiseuille flow simulations by fitting a parabola to the flow profile for the various thermostats.
$l_s$  is the slip length and $\nu_s$ the kinematic viscosity for simulations, where the average phantom-particle momentum is set to zero. The other viscosities are obtain form simulations with a profile-matched phantom-particle momentum.}
\begin{ruledtabular}
\begin{tabular}{l|cccc}
 & $h/\tau$ & 0.1 & 1.0 & 3.0 \\
\hline
 & $\nu /(a^2/\tau)$ & 0.886 & 0.570 & 1.483 \\
 & $\Delta \nu/\%$ & 1.9 & 0.4 & -0.2 \\
LSS & $\nu_s /(a^2/\tau)$ & 0.888 & 0.572 & 1.486 \\
 & $l_s/a$ & 0.167 & 0.054 & 0.079  \\
 & $\Delta \nu_s/\%$ & 2.0 & 0.6 & 0.0 \\ \hline
& $\nu /(a^2/\tau)$ & 0.869 & 0.570 & 1.483  \\
 & $\Delta \nu/\%$ & -0.1 & 0.2 & -0.2 \\
MCS & $\nu_s /(a^2/\tau)$ & 0.869 & 0.571 & 1.482 \\
 & $l_s/a$ &  0.170 &  0.057 & 0.093  \\
 & $\Delta \nu_s/\%$ & -0.1 & 0.4 & -0.3 \\ \hline
 & $\nu /(a^2/\tau)$ & 0.882 & 0.573 & 1.485  \\
 & $\Delta \nu/\%$ & 1.4 & 0.7 & -0.1 \\
MBS & $\nu_s /(a^2/\tau)$ & 0.882 & 0.574 & 1.488 \\
 & $l_s/a$ & 0.172 & 0.057 & 0.090  \\
 & $\Delta \nu_s/\%$ & 1.4 & 0.9 & 0.1 \\
\end{tabular}
\end{ruledtabular}
\end{table}

\subsection{Poiseuille Flow} \label{sec:poiseuille_flow_results}

\begin{figure}[t]
\begin{center}
\includegraphics[width=\columnwidth,angle=0]{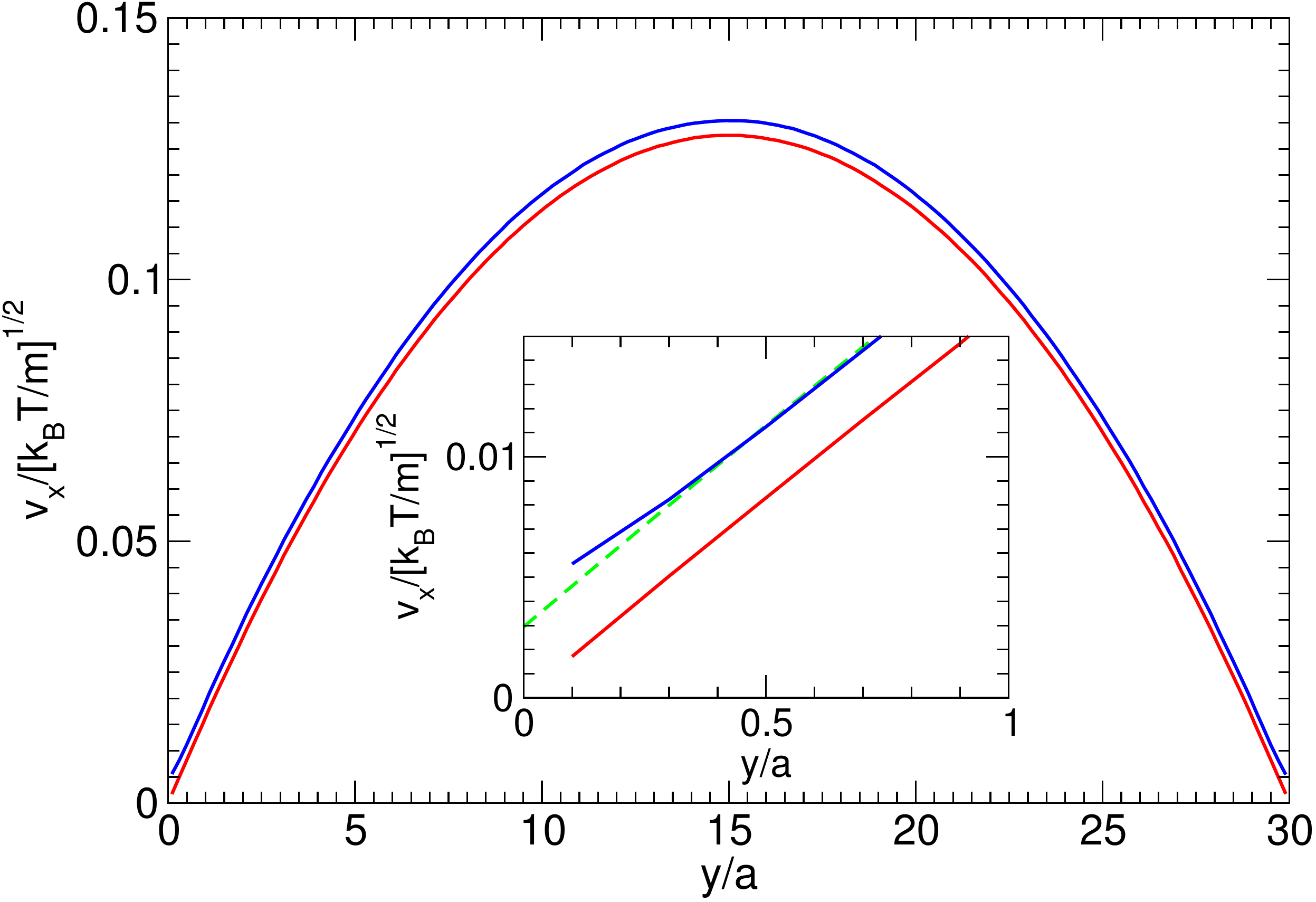}
\caption{(Color online) Velocity profiles of the Poiseuille flow with residual slip (upper curve) and without slip (bottom curve) by assigning a finite (negative) velocity to phantom particles. The time step is $h=0.1 \tau$. The inset shows the profiles close to the solid wall at $y=0$. The dashed line indicates the fit of the velocity profile (\ref{parabolic}) with finite slip length.}
\label{fig:finite_momentum}
\end{center}
\end{figure}

Figure \ref{fig:pois_flow} shows velocity profiles for a force-driven MPC fluid confined between hard walls thermalized by the LSS, MCS, or MBS method, and various collision-time steps. Here, the average momentum of a phantom particle in a wall collision cell is determined by the desired parabolic velocity profile (cf. Sec.~\ref{sec:poiseuille_flow}). As displayed in the figure, this choice yields a zero fluid velocity at the surface (see als Fig.~\ref{fig:finite_momentum}).
Independent of the applied thermostat, the fluid particle temperature and density across the channel are constant. For every collision-time step, we find good agreement between the velocity profiles of the various thermostats. Moreover, the profiles agree well with the parabola with the theoretically determined viscosities. The actually determined viscosities and their deviations from the theoretical values are summarized in Table~\ref{table2}. Here, the simulation data are fitted by the parabola
\begin{align} \label{parabolic}
v_x(y) = \frac{g}{2 \nu}(y+l_s)(L+l_s-y) ,
\end{align}
which yields the slip length $l_s$ and the kinematic viscosity $\nu$.
As expected, we find a zero slip length in simulations where the phantom-particle momentum $\lla {\bm P}\rra \neq 0$.
There are only very minor differences between the viscosities obtained for the various thermostats, and the viscosities themselves agree well with the theoretical values. Thereby, the numerical values are typically somewhat larger, up to approximately $2\%$. The shear-flow simulations show the same trend.

Figure~\ref{fig:finite_momentum} compares velocity profiles for no-slip boundary conditions, where $\lla {\bm P}\rra \neq 0$, with results with residual slip, where we set $\lla {\bm P}\rra =0$ (cf. Sec.~\ref{sec:poiseuille_flow}). There is a finite slip for $\lla {\bm P}\rra =0$,  which implies a shift of the whole velocity profile to larger velocities. A fit by the profile (\ref{parabolic}) yields the slip length $l_s$ and the viscosity $\nu_s$. The respective values are summarized in Table~\ref{table2}. In general, the profiles are excellently fitted by Eq.~(\ref{parabolic}). Despite the differences of the profiles, the viscosities are in close agreement. Since a finite residual slip does not alter the viscosity, a fit with a finite slip length yields a very accurate estimation of the viscosity. However, a fit with zero slip length provides a somewhat different viscosity. Thereby, the overall numerical profile is not very well reproduced by the theoretical parabola. Inclusion of a slip length improves fitting considerably.

\section{Discussion} \label{sec:discussion}

\begin{figure}[t]
\begin{center}
\includegraphics[width=\columnwidth,angle=0]{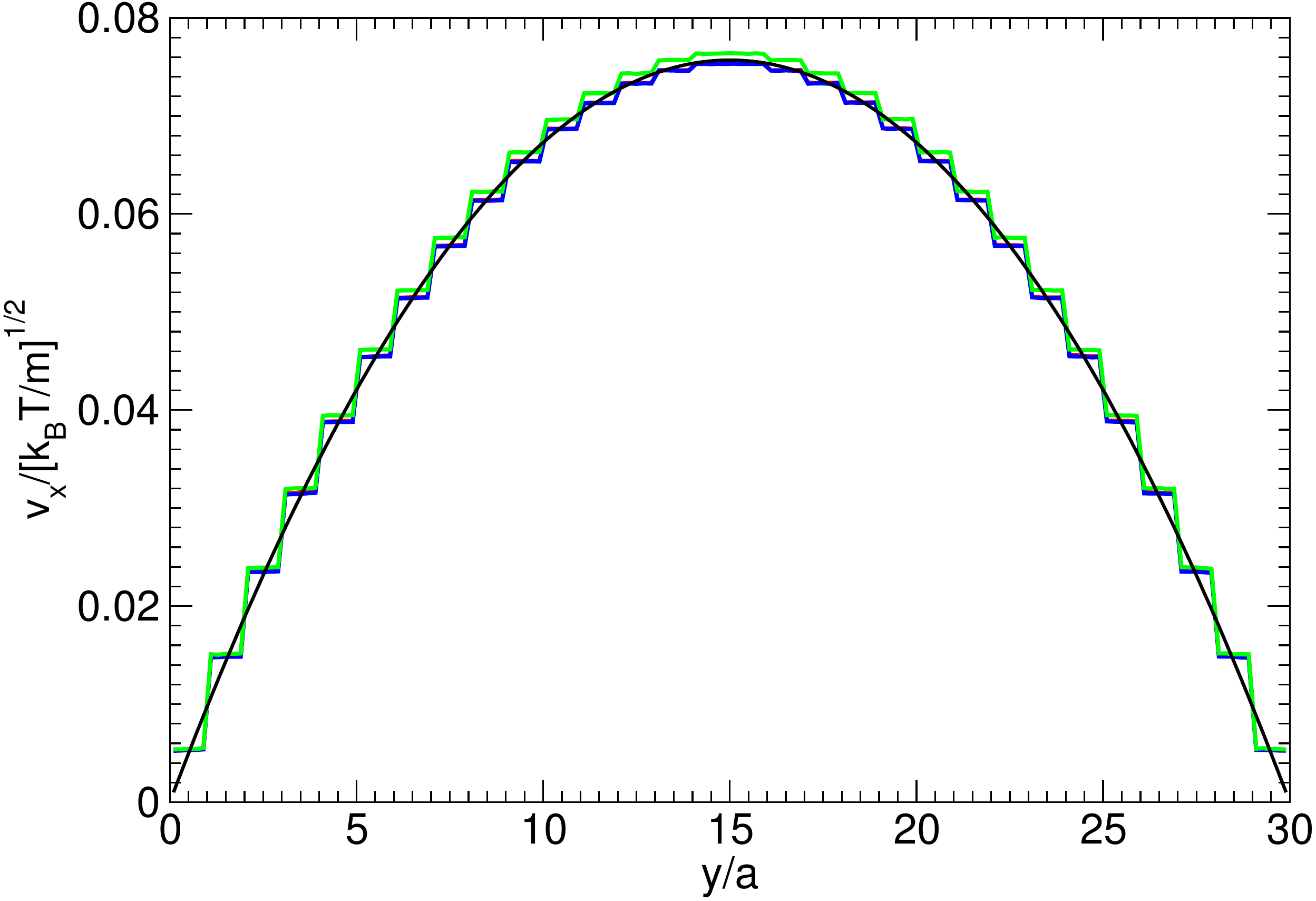}
\caption{(Color online) Velocity profiles for the Monte Carlo scaling (MCS) approach (top, green) and the Maxwell-Boltzmann scaling (MBS) method (bottom, blue) for the collision-time step $h=3.0\tau$. The result for the LSS approach is indistinguishable from the MBS result.  No random shift is applied. The smooth solid line (black) is the theoretical parabolic velocity profile with the analytically determined viscosity presented in Table~\ref{table1}. }
\label{step_pois}
\end{center}
\end{figure}

The viscosity of the MPC fluid is dominated by contributions from collisions at small, and by kinetic contributions (streaming) at large collision-time steps. This suggests that a random shift of the collision lattice can be omitted at large collision-time steps \cite{lamu:01,whit:10,boli:12},
and partially filled collision cells would not matter anymore.
However, lack of a random shift causes various ambiguities. Without random shift and phantom particles, there are only bounce-back interactions during streaming with walls, which  does not prevent slip strictly, because the average velocity at the wall will never be zero during a MPC dynamics step. More severely, the induced velocity profile is no longer smooth on the length scale of a collision cell. As shown in Fig.~\ref{step_pois}, correlations on the cell level lead to essentially constant average velocities of the particles in a collision cell and, hence, to a step-like overall profile. Note that we calculate the velocity profile after a collision. The steps appear for all collision schemes with conserved linear momentum, since the stationary state distribution of the relative velocities $\Delta {\bm v}_i$ is Gaussian with zero mean.  Hence, the average velocity of a particle in a cell after collision is $\lla {\bm v}_i\rra = \lla {\bm v}_{cm}\rra$.  The calculation of the velocities after streaming yields a smoother profile, in particular for very large collision-time steps. If only the velocity of the cell center would be considered, i.e., the bin width for the calculation of the profile is set equal to the size of the collision cell, the steps are invisible and a smooth profile is obtained. As revealed by the in-depth studies of Ref.~\cite{boli:12}, lack or presence of a random shift leads to slightly different velocity profiles, with a higher viscosity in the presence of a random shift. The difference for the studied time step, however, is extremely small and is expected to be even smaller for larger $h$. Importantly, the difference between the velocity profiles is not related to partially filled collision cells, but only to the shift of the collision lattice.   To avoid  ambiguities in the calculation of the velocity profile (after streaming versus after collision),  we recommend to use a random shift of the collision lattice for any time step. This yields a unique viscosity.

\section{Summary and Conclusions} \label{sec:conclusion}

We have presented a detailed evaluation of various thermostating approaches for non-equilibrium multiparticle collision dynamics simulations. The purpose of our studies is twofold. On the one hand, we want to shed light on the accuracy with which the non-equilibrium aspects of the fluid are reproduced or are perturbed by a particular thermostat. On the other hand, we intend to clarify whether there are deviations, and if so, how large,  between the analytically determined viscosity of the MPC fluid and that extracted from simulation. For this purpose, we have determined the fluid viscosities for various MPC collision-time steps by equilibrium simulations via the transverse fluid velocity autocorrelation function and by non-equilibrium simulations calculating the stress tensor under simple shear, as well as the velocity profile for a Poiseuille flow.
In the non-equilibrium simulations, we control temperature on the cell level by three different methods: local simple scaling (LSS), a Monte Carlo-like scheme (MCS) \cite{boli:12}, and by the Maxwell-Boltzmann scaling (MBS) approach  \cite{huan:10.1}.

The calculation of the dynamic structure factor with the equilibrium fluid density fluctuations of the system thermalized by the MCS method yields, at first unexpected,  a pronounced Rayleigh peak for collision-time steps $h/\tau > 1 $. Hence, in such a system thermal fluctuations play a major role and it is closer to an isoentropic rather than an isothermal ensemble. The correct average asymptotic temperature is assumed for many Monte Carlo steps, however, the fluctuations do not correspond to an isothermal ensemble. This implies different transport coefficients compared to an strict isothermal system---they may neither correspond to an adiabatic nor to an isothermal system.

However, our simulations suggest that every employed thermostat leaves the viscosity unchanged, or at least affects it to such small extent that it is difficult to detect by the flow profiles or thermodynamic properties. Hence, we find only minor differences between the viscosities obtained by the various approaches. Thus,  we consider all of them suitable for non-equilibrium simulations for weakly perturbed systems. The drawback of the LSS method, however, is that the velocity distribution of the fluid particles is non-Gaussian, which leads to artifacts in the density and even temperature distribution at larger field strengths. As discussed in Ref.~\cite{huan:10.1}, the MBS method provides accurate results even at large fields.

Looking at the agreement between the analytically predicted viscosities with those determined by simulations, we find slightly larger numerical values in the range $0.2 \le h/\tau < 1$ than theoretically predicted (cf. Table~\ref{table1}). All thermostats yield consistently slightly larger viscosities, with some small variations.

We have only considered the SRD variant of MPC, where fluid velocities are rotated around a randomly oriented axis. In Ref.~\cite{boli:12}, other collision rules have been applied for Poiseuille flow simulations, in particular rotations around the Cartesian axes only. Simulations exploiting the MPC-AT method yield velocity profiles, which agree very well with those determined with the theoretical viscosity for large collision-time steps. However, simulations applying rotations of the relative velocities around one of the randomly selected Cartesian axis yields considerable deviations between simulation and theoretical results. Applying the same rule, we also find larger deviations than those found by the above applied collision rule. Hence, the collision rule affects the fluid behavior under non-equilibrium conditions. In the axis-rotation scheme, there seem to be considerable correlations of the fluid particles in a collision cell, more than in the other algorithms.

Our simulations reveal a major effect of the violation of Galilean invariance on the flow properties  in the form of stair-like profiles, for both, simple shear and Poiseuille flow.  The  effect as such  is independent of the collision-time step. Signatures of such steps have also been reported in Ref.~\cite{boli:12}. As we have shown, the steps completely disappear when a random shift of the collision lattice is applied. Thus, we strongly recommend to apply a random shift of the collision lattice even for large collision-time steps, although physically relevant fluid properties can only be expected on length scales larger than a collision cell.

Moreover, the random shift is intimately  connected with the boundary condition. A no-slip boundary condition is best fulfilled by applying a random shift and inclusion of phantom particles \cite{lamu:01,whit:10}, for both, stationary surfaces as well as dissolved solid bodies \cite{goet:07,pobl:14}.
Further investigations of the boundary conditions on the dynamics of colloids are currently under way, with emphasis on the differences in colloid dynamics between slip and no-slip boundaries.

\appendix
\section{Transport Coefficients} \label{appendix}

Here, the various transport coefficients defined in Section \ref{sec:hydro_fluct} are
given in terms of the MPC-SRD fluid parameters \cite{tuez:06,huan:12}. We assume that $\lla N_c \rra \gg 1$, such that $\lla N_c \rra - 1 + e^{- \lla N_c \rra} \approx \lla N_c \rra - 1$.

The thermal diffusion coefficient $D_T$  is given by
\begin{align}
D_T & = D_T^c + D_T^k,
\end{align}
with
\begin{align}
D_T^c &  =  \frac{a^2}{15 h \lla N_c \rra} \left( 1-\frac{1}{\lla N_c \rra}\right) \,
(1-\cos \alpha) \ , \\
D_T^k & = \frac{k_B Th}{2 m}
\left[
\frac{3}{1-\cos \alpha} - 1 + \frac{6}{\lla N_c \rra} \left(
\frac{4}{5}-\frac{1}{4}\frac{1}{\sin^2 \alpha/2}\right)
\right].
\end{align}
The specific heat capacities are
\begin{equation}
c_v = \frac{3 k_B}{2 m} \ , \qquad c_p = c_v + \frac{k_B}{m} \ , \qquad
\gamma =  \frac{c_p}{c_v} = \frac{5}{3}.
\end{equation}
The sound attenuation factor of an adiabatic system is defined as
\begin{align}
\Gamma_s & = \frac{1}{2} \left[D_T (\gamma - 1) + \tilde \nu \right] ,
\end{align}
with
\begin{align}
\tilde \nu & = \frac{4}{3}  \nu^k + \nu^c = \frac{4}{3} \nu - \frac{1}{3} \nu^c,
\end{align}
and the kinematic viscosity $\nu = \eta/\varrho = \nu^k + \nu^c$. The viscosities $\eta^c$ and $\eta^k$ are defined in Eqs.~(\ref{visc_coll}) and (\ref{visc_kin}).

For an isothermal fluid, the sound attenuation factor is
\begin{align}
\Gamma & = \frac{1}{2} \tilde \nu  .
\end{align}


\begin{thebibliography}{97}%
\makeatletter
\providecommand \@ifxundefined [1]{%
 \@ifx{#1\undefined}
}%
\providecommand \@ifnum [1]{%
 \ifnum #1\expandafter \@firstoftwo
 \else \expandafter \@secondoftwo
 \fi
}%
\providecommand \@ifx [1]{%
 \ifx #1\expandafter \@firstoftwo
 \else \expandafter \@secondoftwo
 \fi
}%
\providecommand \natexlab [1]{#1}%
\providecommand \enquote  [1]{``#1''}%
\providecommand \bibnamefont  [1]{#1}%
\providecommand \bibfnamefont [1]{#1}%
\providecommand \citenamefont [1]{#1}%
\providecommand \href@noop [0]{\@secondoftwo}%
\providecommand \href [0]{\begingroup \@sanitize@url \@href}%
\providecommand \@href[1]{\@@startlink{#1}\@@href}%
\providecommand \@@href[1]{\endgroup#1\@@endlink}%
\providecommand \@sanitize@url [0]{\catcode `\\12\catcode `\$12\catcode
  `\&12\catcode `\#12\catcode `\^12\catcode `\_12\catcode `\%12\relax}%
\providecommand \@@startlink[1]{}%
\providecommand \@@endlink[0]{}%
\providecommand \url  [0]{\begingroup\@sanitize@url \@url }%
\providecommand \@url [1]{\endgroup\@href {#1}{\urlprefix }}%
\providecommand \urlprefix  [0]{URL }%
\providecommand \Eprint [0]{\href }%
\providecommand \doibase [0]{http://dx.doi.org/}%
\providecommand \selectlanguage [0]{\@gobble}%
\providecommand \bibinfo  [0]{\@secondoftwo}%
\providecommand \bibfield  [0]{\@secondoftwo}%
\providecommand \translation [1]{[#1]}%
\providecommand \BibitemOpen [0]{}%
\providecommand \bibitemStop [0]{}%
\providecommand \bibitemNoStop [0]{.\EOS\space}%
\providecommand \EOS [0]{\spacefactor3000\relax}%
\providecommand \BibitemShut  [1]{\csname bibitem#1\endcsname}%
\let\auto@bib@innerbib\@empty
\bibitem [{\citenamefont {Malevanets}\ and\ \citenamefont
  {Kapral}(1999)}]{male:99}%
  \BibitemOpen
  \bibfield  {author} {\bibinfo {author} {\bibfnamefont {A.}~\bibnamefont
  {Malevanets}}\ and\ \bibinfo {author} {\bibfnamefont {R.}~\bibnamefont
  {Kapral}},\ }\href@noop {} {\bibfield  {journal} {\bibinfo  {journal} {J.
  Chem. Phys.}\ }\textbf {\bibinfo {volume} {110}},\ \bibinfo {pages} {8605}
  (\bibinfo {year} {1999})}\BibitemShut {NoStop}%
\bibitem [{\citenamefont {Malevanets}\ and\ \citenamefont
  {Kapral}(2000)}]{male:00}%
  \BibitemOpen
  \bibfield  {author} {\bibinfo {author} {\bibfnamefont {A.}~\bibnamefont
  {Malevanets}}\ and\ \bibinfo {author} {\bibfnamefont {R.}~\bibnamefont
  {Kapral}},\ }\href@noop {} {\bibfield  {journal} {\bibinfo  {journal} {J.
  Chem. Phys.}\ }\textbf {\bibinfo {volume} {112}},\ \bibinfo {pages} {7260}
  (\bibinfo {year} {2000})}\BibitemShut {NoStop}%
\bibitem [{\citenamefont {Kapral}(2008)}]{kapr:08}%
  \BibitemOpen
  \bibfield  {author} {\bibinfo {author} {\bibfnamefont {R.}~\bibnamefont
  {Kapral}},\ }\href@noop {} {\bibfield  {journal} {\bibinfo  {journal} {Adv.
  Chem. Phys.}\ }\textbf {\bibinfo {volume} {140}},\ \bibinfo {pages} {89}
  (\bibinfo {year} {2008})}\BibitemShut {NoStop}%
\bibitem [{\citenamefont {Gompper}\ \emph {et~al.}(2009)\citenamefont
  {Gompper}, \citenamefont {Ihle}, \citenamefont {Kroll},\ and\ \citenamefont
  {Winkler}}]{gomp:09}%
  \BibitemOpen
  \bibfield  {author} {\bibinfo {author} {\bibfnamefont {G.}~\bibnamefont
  {Gompper}}, \bibinfo {author} {\bibfnamefont {T.}~\bibnamefont {Ihle}},
  \bibinfo {author} {\bibfnamefont {D.~M.}\ \bibnamefont {Kroll}}, \ and\
  \bibinfo {author} {\bibfnamefont {R.~G.}\ \bibnamefont {Winkler}},\ }\href
  {\doibase 10.1007/978-3-540-87706-6{\_1}} {\bibfield  {journal} {\bibinfo
  {journal} {Adv. Polym. Sci.}\ }\textbf {\bibinfo {volume} {221}},\ \bibinfo
  {pages} {1} (\bibinfo {year} {2009})}\BibitemShut {NoStop}%
\bibitem [{\citenamefont {Lee}\ and\ \citenamefont {Kapral}(2004)}]{lee:04}%
  \BibitemOpen
  \bibfield  {author} {\bibinfo {author} {\bibfnamefont {S.~H.}\ \bibnamefont
  {Lee}}\ and\ \bibinfo {author} {\bibfnamefont {R.}~\bibnamefont {Kapral}},\
  }\href {\doibase 10.1063/1.1815291} {\bibfield  {journal} {\bibinfo
  {journal} {J. Chem. Phys.}\ }\textbf {\bibinfo {volume} {121}},\ \bibinfo
  {pages} {11163} (\bibinfo {year} {2004})}\BibitemShut {NoStop}%
\bibitem [{\citenamefont {Hecht}\ \emph {et~al.}(2005)\citenamefont {Hecht},
  \citenamefont {Harting}, \citenamefont {Ihle},\ and\ \citenamefont
  {Herrmann}}]{hech:05}%
  \BibitemOpen
  \bibfield  {author} {\bibinfo {author} {\bibfnamefont {M.}~\bibnamefont
  {Hecht}}, \bibinfo {author} {\bibfnamefont {J.}~\bibnamefont {Harting}},
  \bibinfo {author} {\bibfnamefont {T.}~\bibnamefont {Ihle}}, \ and\ \bibinfo
  {author} {\bibfnamefont {H.~J.}\ \bibnamefont {Herrmann}},\ }\href {\doibase
  10.1103/PhysRevE.72.011408} {\bibfield  {journal} {\bibinfo  {journal} {Phys.
  Rev. E}\ }\textbf {\bibinfo {volume} {72}},\ \bibinfo {pages} {011408}
  (\bibinfo {year} {2005})}\BibitemShut {NoStop}%
\bibitem [{\citenamefont {Padding}\ and\ \citenamefont
  {Louis}(2006)}]{padd:06}%
  \BibitemOpen
  \bibfield  {author} {\bibinfo {author} {\bibfnamefont {J.~T.}\ \bibnamefont
  {Padding}}\ and\ \bibinfo {author} {\bibfnamefont {A.~A.}\ \bibnamefont
  {Louis}},\ }\href@noop {} {\bibfield  {journal} {\bibinfo  {journal} {Phys.
  Rev. E}\ }\textbf {\bibinfo {volume} {74}},\ \bibinfo {pages} {031402}
  (\bibinfo {year} {2006})}\BibitemShut {NoStop}%
\bibitem [{\citenamefont {G{\"o}tze}\ \emph {et~al.}(2007)\citenamefont
  {G{\"o}tze}, \citenamefont {Noguchi},\ and\ \citenamefont
  {Gompper}}]{goet:07}%
  \BibitemOpen
  \bibfield  {author} {\bibinfo {author} {\bibfnamefont {I.~O.}\ \bibnamefont
  {G{\"o}tze}}, \bibinfo {author} {\bibfnamefont {H.}~\bibnamefont {Noguchi}},
  \ and\ \bibinfo {author} {\bibfnamefont {G.}~\bibnamefont {Gompper}},\
  }\href@noop {} {\bibfield  {journal} {\bibinfo  {journal} {Phys. Rev. E}\
  }\textbf {\bibinfo {volume} {76}},\ \bibinfo {pages} {046705} (\bibinfo
  {year} {2007})}\BibitemShut {NoStop}%
\bibitem [{\citenamefont {Petersen}\ \emph {et~al.}(2010)\citenamefont
  {Petersen}, \citenamefont {Lechman}, \citenamefont {Plimpton}, \citenamefont
  {G.~S.~Grest},\ and\ \citenamefont {Schunk}}]{pete:10}%
  \BibitemOpen
  \bibfield  {author} {\bibinfo {author} {\bibfnamefont {M.~K.}\ \bibnamefont
  {Petersen}}, \bibinfo {author} {\bibfnamefont {J.~B.}\ \bibnamefont
  {Lechman}}, \bibinfo {author} {\bibfnamefont {S.~J.}\ \bibnamefont
  {Plimpton}}, \bibinfo {author} {\bibfnamefont {P.~J. i.~Â.}\ \bibnamefont
  {G.~S.~Grest}}, \ and\ \bibinfo {author} {\bibfnamefont {P.~R.}\ \bibnamefont
  {Schunk}},\ }\href@noop {} {\bibfield  {journal} {\bibinfo  {journal} {J.
  Chem. Phys.}\ }\textbf {\bibinfo {volume} {132}},\ \bibinfo {pages} {174106}
  (\bibinfo {year} {2010})}\BibitemShut {NoStop}%
\bibitem [{\citenamefont {Whitmer}\ and\ \citenamefont
  {Luijten}(2010)}]{whit:10}%
  \BibitemOpen
  \bibfield  {author} {\bibinfo {author} {\bibfnamefont {J.~K.}\ \bibnamefont
  {Whitmer}}\ and\ \bibinfo {author} {\bibfnamefont {E.}~\bibnamefont
  {Luijten}},\ }\href {\doibase doi:10.1088/0953-8984/22/10/104106} {\bibfield
  {journal} {\bibinfo  {journal} {J. Phys.: Condens. Matter}\ }\textbf
  {\bibinfo {volume} {22}},\ \bibinfo {pages} {104106} (\bibinfo {year}
  {2010})}\BibitemShut {NoStop}%
\bibitem [{\citenamefont {Franosch}\ \emph {et~al.}(2011)\citenamefont
  {Franosch}, \citenamefont {Grimm}, \citenamefont {Belushkin}, \citenamefont
  {Mor}, \citenamefont {Foffi}, \citenamefont {Forr{\'{o}}},\ and\
  \citenamefont {Jeney}}]{fran:11}%
  \BibitemOpen
  \bibfield  {author} {\bibinfo {author} {\bibfnamefont {T.}~\bibnamefont
  {Franosch}}, \bibinfo {author} {\bibfnamefont {M.}~\bibnamefont {Grimm}},
  \bibinfo {author} {\bibfnamefont {M.}~\bibnamefont {Belushkin}}, \bibinfo
  {author} {\bibfnamefont {F.~M.}\ \bibnamefont {Mor}}, \bibinfo {author}
  {\bibfnamefont {G.}~\bibnamefont {Foffi}}, \bibinfo {author} {\bibfnamefont
  {L.}~\bibnamefont {Forr{\'{o}}}}, \ and\ \bibinfo {author} {\bibfnamefont
  {S.}~\bibnamefont {Jeney}},\ }\href@noop {} {\bibfield  {journal} {\bibinfo
  {journal} {Nature}\ }\textbf {\bibinfo {volume} {478}},\ \bibinfo {pages}
  {85} (\bibinfo {year} {2011})}\BibitemShut {NoStop}%
\bibitem [{\citenamefont {Belushkin}\ \emph {et~al.}(2011)\citenamefont
  {Belushkin}, \citenamefont {Winkler},\ and\ \citenamefont {Foffi}}]{belu:11}%
  \BibitemOpen
  \bibfield  {author} {\bibinfo {author} {\bibfnamefont {M.}~\bibnamefont
  {Belushkin}}, \bibinfo {author} {\bibfnamefont {R.~G.}\ \bibnamefont
  {Winkler}}, \ and\ \bibinfo {author} {\bibfnamefont {G.}~\bibnamefont
  {Foffi}},\ }\href@noop {} {\bibfield  {journal} {\bibinfo  {journal} {J.
  Phys. Chem. B.}\ }\textbf {\bibinfo {volume} {115}},\ \bibinfo {pages}
  {14263} (\bibinfo {year} {2011})}\BibitemShut {NoStop}%
\bibitem [{\citenamefont {Malevanets}\ and\ \citenamefont
  {Yeomans}(2000)}]{male:00.1}%
  \BibitemOpen
  \bibfield  {author} {\bibinfo {author} {\bibfnamefont {A.}~\bibnamefont
  {Malevanets}}\ and\ \bibinfo {author} {\bibfnamefont {J.~M.}\ \bibnamefont
  {Yeomans}},\ }\href@noop {} {\bibfield  {journal} {\bibinfo  {journal}
  {Europhys. Lett.}\ }\textbf {\bibinfo {volume} {52}},\ \bibinfo {pages} {231}
  (\bibinfo {year} {2000})}\BibitemShut {NoStop}%
\bibitem [{\citenamefont {Ripoll}\ \emph {et~al.}(2004)\citenamefont {Ripoll},
  \citenamefont {Mussawisade}, \citenamefont {Winkler},\ and\ \citenamefont
  {Gompper}}]{ripo:04}%
  \BibitemOpen
  \bibfield  {author} {\bibinfo {author} {\bibfnamefont {M.}~\bibnamefont
  {Ripoll}}, \bibinfo {author} {\bibfnamefont {K.}~\bibnamefont {Mussawisade}},
  \bibinfo {author} {\bibfnamefont {R.~G.}\ \bibnamefont {Winkler}}, \ and\
  \bibinfo {author} {\bibfnamefont {G.}~\bibnamefont {Gompper}},\ }\href@noop
  {} {\bibfield  {journal} {\bibinfo  {journal} {Europhys. Lett.}\ }\textbf
  {\bibinfo {volume} {68}},\ \bibinfo {pages} {106} (\bibinfo {year}
  {2004})}\BibitemShut {NoStop}%
\bibitem [{\citenamefont {Mussawisade}\ \emph {et~al.}(2005)\citenamefont
  {Mussawisade}, \citenamefont {Ripoll}, \citenamefont {Winkler},\ and\
  \citenamefont {Gompper}}]{muss:05}%
  \BibitemOpen
  \bibfield  {author} {\bibinfo {author} {\bibfnamefont {K.}~\bibnamefont
  {Mussawisade}}, \bibinfo {author} {\bibfnamefont {M.}~\bibnamefont {Ripoll}},
  \bibinfo {author} {\bibfnamefont {R.~G.}\ \bibnamefont {Winkler}}, \ and\
  \bibinfo {author} {\bibfnamefont {G.}~\bibnamefont {Gompper}},\ }\href
  {\doibase 10.1063/1.2041527} {\bibfield  {journal} {\bibinfo  {journal} {J.
  Chem. Phys.}\ }\textbf {\bibinfo {volume} {123}},\ \bibinfo {pages} {144905}
  (\bibinfo {year} {2005})}\BibitemShut {NoStop}%
\bibitem [{\citenamefont {Huang}\ \emph
  {et~al.}(2010{\natexlab{a}})\citenamefont {Huang}, \citenamefont {Winkler},
  \citenamefont {Sutmann},\ and\ \citenamefont {Gompper}}]{huan:10}%
  \BibitemOpen
  \bibfield  {author} {\bibinfo {author} {\bibfnamefont {C.-C.}\ \bibnamefont
  {Huang}}, \bibinfo {author} {\bibfnamefont {R.~G.}\ \bibnamefont {Winkler}},
  \bibinfo {author} {\bibfnamefont {G.}~\bibnamefont {Sutmann}}, \ and\
  \bibinfo {author} {\bibfnamefont {G.}~\bibnamefont {Gompper}},\ }\href
  {\doibase 10.1021/ma101836x} {\bibfield  {journal} {\bibinfo  {journal}
  {Macromolecules}\ }\textbf {\bibinfo {volume} {43}},\ \bibinfo {pages}
  {10107} (\bibinfo {year} {2010}{\natexlab{a}})}\BibitemShut {NoStop}%
\bibitem [{\citenamefont {Huang}\ \emph {et~al.}(2013)\citenamefont {Huang},
  \citenamefont {Gompper},\ and\ \citenamefont {Winkler}}]{huan:13}%
  \BibitemOpen
  \bibfield  {author} {\bibinfo {author} {\bibfnamefont {C.~C.}\ \bibnamefont
  {Huang}}, \bibinfo {author} {\bibfnamefont {G.}~\bibnamefont {Gompper}}, \
  and\ \bibinfo {author} {\bibfnamefont {R.~G.}\ \bibnamefont {Winkler}},\
  }\href {\doibase 10.1063/1.4799877} {\bibfield  {journal} {\bibinfo
  {journal} {J. Chem. Phys.}\ }\textbf {\bibinfo {volume} {138}},\ \bibinfo
  {pages} {144902} (\bibinfo {year} {2013})}\BibitemShut {NoStop}%
\bibitem [{\citenamefont {Lamura}\ \emph {et~al.}(2001)\citenamefont {Lamura},
  \citenamefont {Gompper}, \citenamefont {Ihle},\ and\ \citenamefont
  {Kroll}}]{lamu:01}%
  \BibitemOpen
  \bibfield  {author} {\bibinfo {author} {\bibfnamefont {A.}~\bibnamefont
  {Lamura}}, \bibinfo {author} {\bibfnamefont {G.}~\bibnamefont {Gompper}},
  \bibinfo {author} {\bibfnamefont {T.}~\bibnamefont {Ihle}}, \ and\ \bibinfo
  {author} {\bibfnamefont {D.~M.}\ \bibnamefont {Kroll}},\ }\href@noop {}
  {\bibfield  {journal} {\bibinfo  {journal} {Europhys. Lett.}\ }\textbf
  {\bibinfo {volume} {56}},\ \bibinfo {pages} {319} (\bibinfo {year}
  {2001})}\BibitemShut {NoStop}%
\bibitem [{\citenamefont {Allahyarov}\ and\ \citenamefont
  {Gompper}(2002)}]{alla:02}%
  \BibitemOpen
  \bibfield  {author} {\bibinfo {author} {\bibfnamefont {E.}~\bibnamefont
  {Allahyarov}}\ and\ \bibinfo {author} {\bibfnamefont {G.}~\bibnamefont
  {Gompper}},\ }\href@noop {} {\bibfield  {journal} {\bibinfo  {journal} {Phys.
  Rev. E}\ }\textbf {\bibinfo {volume} {66}},\ \bibinfo {pages} {036702}
  (\bibinfo {year} {2002})}\BibitemShut {NoStop}%
\bibitem [{\citenamefont {Winkler}\ \emph {et~al.}(2004)\citenamefont
  {Winkler}, \citenamefont {Mussawisade}, \citenamefont {Ripoll},\ and\
  \citenamefont {Gompper}}]{wink:04}%
  \BibitemOpen
  \bibfield  {author} {\bibinfo {author} {\bibfnamefont {R.~G.}\ \bibnamefont
  {Winkler}}, \bibinfo {author} {\bibfnamefont {K.}~\bibnamefont
  {Mussawisade}}, \bibinfo {author} {\bibfnamefont {M.}~\bibnamefont {Ripoll}},
  \ and\ \bibinfo {author} {\bibfnamefont {G.}~\bibnamefont {Gompper}},\
  }\href@noop {} {\bibfield  {journal} {\bibinfo  {journal} {J. Phys.: Condens.
  Matter}\ }\textbf {\bibinfo {volume} {16}},\ \bibinfo {pages} {S3941}
  (\bibinfo {year} {2004})}\BibitemShut {NoStop}%
\bibitem [{\citenamefont {Padding}\ and\ \citenamefont
  {Louis}(2004)}]{padd:04}%
  \BibitemOpen
  \bibfield  {author} {\bibinfo {author} {\bibfnamefont {J.~T.}\ \bibnamefont
  {Padding}}\ and\ \bibinfo {author} {\bibfnamefont {A.~A.}\ \bibnamefont
  {Louis}},\ }\href@noop {} {\bibfield  {journal} {\bibinfo  {journal} {Phys.
  Rev. Lett.}\ }\textbf {\bibinfo {volume} {93}},\ \bibinfo {pages} {220601}
  (\bibinfo {year} {2004})}\BibitemShut {NoStop}%
\bibitem [{\citenamefont {Ripoll}\ \emph {et~al.}(2008)\citenamefont {Ripoll},
  \citenamefont {Holmqvist}, \citenamefont {Winkler}, \citenamefont {Gompper},
  \citenamefont {Dhont},\ and\ \citenamefont {Lettinga}}]{ripo:08}%
  \BibitemOpen
  \bibfield  {author} {\bibinfo {author} {\bibfnamefont {M.}~\bibnamefont
  {Ripoll}}, \bibinfo {author} {\bibfnamefont {P.}~\bibnamefont {Holmqvist}},
  \bibinfo {author} {\bibfnamefont {R.~G.}\ \bibnamefont {Winkler}}, \bibinfo
  {author} {\bibfnamefont {G.}~\bibnamefont {Gompper}}, \bibinfo {author}
  {\bibfnamefont {J.~K.~G.}\ \bibnamefont {Dhont}}, \ and\ \bibinfo {author}
  {\bibfnamefont {M.~P.}\ \bibnamefont {Lettinga}},\ }\href@noop {} {\bibfield
  {journal} {\bibinfo  {journal} {Phys. Rev. Lett.}\ }\textbf {\bibinfo
  {volume} {101}},\ \bibinfo {pages} {168302} (\bibinfo {year}
  {2008})}\BibitemShut {NoStop}%
\bibitem [{\citenamefont {Wysocki}\ \emph {et~al.}(2009)\citenamefont
  {Wysocki}, \citenamefont {Royall}, \citenamefont {Winkler}, \citenamefont
  {Gompper}, \citenamefont {Tanaka}, \citenamefont {van Blaaderen},\ and\
  \citenamefont {L{\"o}wen}}]{wyso:09}%
  \BibitemOpen
  \bibfield  {author} {\bibinfo {author} {\bibfnamefont {A.}~\bibnamefont
  {Wysocki}}, \bibinfo {author} {\bibfnamefont {C.~P.}\ \bibnamefont {Royall}},
  \bibinfo {author} {\bibfnamefont {R.~G.}\ \bibnamefont {Winkler}}, \bibinfo
  {author} {\bibfnamefont {G.}~\bibnamefont {Gompper}}, \bibinfo {author}
  {\bibfnamefont {H.}~\bibnamefont {Tanaka}}, \bibinfo {author} {\bibfnamefont
  {A.}~\bibnamefont {van Blaaderen}}, \ and\ \bibinfo {author} {\bibfnamefont
  {H.}~\bibnamefont {L{\"o}wen}},\ }\href@noop {} {\bibfield  {journal}
  {\bibinfo  {journal} {Soft Matter}\ }\textbf {\bibinfo {volume} {5}},\
  \bibinfo {pages} {1340} (\bibinfo {year} {2009})}\BibitemShut {NoStop}%
\bibitem [{\citenamefont {G{\"o}tze}\ and\ \citenamefont
  {Gompper}(2010{\natexlab{a}})}]{goet:10.1}%
  \BibitemOpen
  \bibfield  {author} {\bibinfo {author} {\bibfnamefont {I.~O.}\ \bibnamefont
  {G{\"o}tze}}\ and\ \bibinfo {author} {\bibfnamefont {G.}~\bibnamefont
  {Gompper}},\ }\href@noop {} {\bibfield  {journal} {\bibinfo  {journal} {EPL}\
  }\textbf {\bibinfo {volume} {92}},\ \bibinfo {pages} {64003} (\bibinfo {year}
  {2010}{\natexlab{a}})}\BibitemShut {NoStop}%
\bibitem [{\citenamefont {Singh}\ \emph {et~al.}(2011)\citenamefont {Singh},
  \citenamefont {Winkler},\ and\ \citenamefont {Gompper}}]{sing:11}%
  \BibitemOpen
  \bibfield  {author} {\bibinfo {author} {\bibfnamefont {S.~P.}\ \bibnamefont
  {Singh}}, \bibinfo {author} {\bibfnamefont {R.~G.}\ \bibnamefont {Winkler}},
  \ and\ \bibinfo {author} {\bibfnamefont {G.}~\bibnamefont {Gompper}},\
  }\href@noop {} {\bibfield  {journal} {\bibinfo  {journal} {Phys. Rev. Lett.}\
  }\textbf {\bibinfo {volume} {107}},\ \bibinfo {pages} {158301} (\bibinfo
  {year} {2011})}\BibitemShut {NoStop}%
\bibitem [{\citenamefont {Webster}\ and\ \citenamefont
  {Yeomans}(2005)}]{webs:05}%
  \BibitemOpen
  \bibfield  {author} {\bibinfo {author} {\bibfnamefont {M.~A.}\ \bibnamefont
  {Webster}}\ and\ \bibinfo {author} {\bibfnamefont {J.~M.}\ \bibnamefont
  {Yeomans}},\ }\href@noop {} {\bibfield  {journal} {\bibinfo  {journal} {J.
  Chem. Phys.}\ }\textbf {\bibinfo {volume} {122}},\ \bibinfo {pages} {164903}
  (\bibinfo {year} {2005})}\BibitemShut {NoStop}%
\bibitem [{\citenamefont {Ryder}\ and\ \citenamefont
  {Yeomans}(2006)}]{ryde:06}%
  \BibitemOpen
  \bibfield  {author} {\bibinfo {author} {\bibfnamefont {J.~F.}\ \bibnamefont
  {Ryder}}\ and\ \bibinfo {author} {\bibfnamefont {J.~M.}\ \bibnamefont
  {Yeomans}},\ }\href@noop {} {\bibfield  {journal} {\bibinfo  {journal} {J.
  Chem. Phys.}\ }\textbf {\bibinfo {volume} {125}},\ \bibinfo {pages} {194906}
  (\bibinfo {year} {2006})}\BibitemShut {NoStop}%
\bibitem [{\citenamefont {Ripoll}\ \emph {et~al.}(2006)\citenamefont {Ripoll},
  \citenamefont {Winkler},\ and\ \citenamefont {Gompper}}]{ripo:06}%
  \BibitemOpen
  \bibfield  {author} {\bibinfo {author} {\bibfnamefont {M.}~\bibnamefont
  {Ripoll}}, \bibinfo {author} {\bibfnamefont {R.~G.}\ \bibnamefont {Winkler}},
  \ and\ \bibinfo {author} {\bibfnamefont {G.}~\bibnamefont {Gompper}},\ }\href
  {\doibase 10.1103/PhysRevLett.96.188302} {\bibfield  {journal} {\bibinfo
  {journal} {Phys. Rev. Lett.}\ }\textbf {\bibinfo {volume} {96}},\ \bibinfo
  {pages} {188302} (\bibinfo {year} {2006})}\BibitemShut {NoStop}%
\bibitem [{\citenamefont {Cannavacciuolo}\ \emph {et~al.}(2008)\citenamefont
  {Cannavacciuolo}, \citenamefont {Winkler},\ and\ \citenamefont
  {Gompper}}]{cann:08}%
  \BibitemOpen
  \bibfield  {author} {\bibinfo {author} {\bibfnamefont {L.}~\bibnamefont
  {Cannavacciuolo}}, \bibinfo {author} {\bibfnamefont {R.~G.}\ \bibnamefont
  {Winkler}}, \ and\ \bibinfo {author} {\bibfnamefont {G.}~\bibnamefont
  {Gompper}},\ }\href@noop {} {\bibfield  {journal} {\bibinfo  {journal} {EPL}\
  }\textbf {\bibinfo {volume} {83}},\ \bibinfo {pages} {34007} (\bibinfo {year}
  {2008})}\BibitemShut {NoStop}%
\bibitem [{\citenamefont {Frank}\ and\ \citenamefont
  {Winkler}(2008)}]{fran:08}%
  \BibitemOpen
  \bibfield  {author} {\bibinfo {author} {\bibfnamefont {S.}~\bibnamefont
  {Frank}}\ and\ \bibinfo {author} {\bibfnamefont {R.~G.}\ \bibnamefont
  {Winkler}},\ }\href {\doibase 10.1209/0295-5075/83/38004} {\bibfield
  {journal} {\bibinfo  {journal} {Europhys. Lett.}\ }\textbf {\bibinfo {volume}
  {83}},\ \bibinfo {pages} {38004} (\bibinfo {year} {2008})}\BibitemShut
  {NoStop}%
\bibitem [{\citenamefont {Nikoubashman}\ and\ \citenamefont
  {Likos}(2010)}]{niko:10}%
  \BibitemOpen
  \bibfield  {author} {\bibinfo {author} {\bibfnamefont {A.}~\bibnamefont
  {Nikoubashman}}\ and\ \bibinfo {author} {\bibfnamefont {C.~N.}\ \bibnamefont
  {Likos}},\ }\href@noop {} {\bibfield  {journal} {\bibinfo  {journal} {J.
  Chem. Phys.}\ }\textbf {\bibinfo {volume} {133}},\ \bibinfo {pages} {074901}
  (\bibinfo {year} {2010})}\BibitemShut {NoStop}%
\bibitem [{\citenamefont {Fedosov}\ \emph {et~al.}(2012)\citenamefont
  {Fedosov}, \citenamefont {Singh}, \citenamefont {Chatterji}, \citenamefont
  {Winkler},\ and\ \citenamefont {Gompper}}]{fedo:12}%
  \BibitemOpen
  \bibfield  {author} {\bibinfo {author} {\bibfnamefont {D.~A.}\ \bibnamefont
  {Fedosov}}, \bibinfo {author} {\bibfnamefont {S.~P.}\ \bibnamefont {Singh}},
  \bibinfo {author} {\bibfnamefont {A.}~\bibnamefont {Chatterji}}, \bibinfo
  {author} {\bibfnamefont {R.~G.}\ \bibnamefont {Winkler}}, \ and\ \bibinfo
  {author} {\bibfnamefont {G.}~\bibnamefont {Gompper}},\ }\href {\doibase
  10.1039/c2sm07009j} {\bibfield  {journal} {\bibinfo  {journal} {Soft Matter}\
  }\textbf {\bibinfo {volume} {8}},\ \bibinfo {pages} {4109} (\bibinfo {year}
  {2012})}\BibitemShut {NoStop}%
\bibitem [{\citenamefont {Huang}\ \emph
  {et~al.}(2012{\natexlab{a}})\citenamefont {Huang}, \citenamefont {Chen},\
  and\ \citenamefont {Mikhailov}}]{huan:12.1}%
  \BibitemOpen
  \bibfield  {author} {\bibinfo {author} {\bibfnamefont {M.-J.}\ \bibnamefont
  {Huang}}, \bibinfo {author} {\bibfnamefont {H.-Y.}\ \bibnamefont {Chen}}, \
  and\ \bibinfo {author} {\bibfnamefont {A.}~\bibnamefont {Mikhailov}},\ }\href
  {\doibase 10.1140/epje/i2012-12119-5} {\bibfield  {journal} {\bibinfo
  {journal} {Eur. Phys. J. E}\ }\textbf {\bibinfo {volume} {35}},\ \bibinfo
  {pages} {119} (\bibinfo {year} {2012}{\natexlab{a}})}\BibitemShut {NoStop}%
\bibitem [{\citenamefont {Chelakkot}\ \emph {et~al.}(2012)\citenamefont
  {Chelakkot}, \citenamefont {Winkler},\ and\ \citenamefont
  {Gompper}}]{chel:12}%
  \BibitemOpen
  \bibfield  {author} {\bibinfo {author} {\bibfnamefont {R.}~\bibnamefont
  {Chelakkot}}, \bibinfo {author} {\bibfnamefont {R.~G.}\ \bibnamefont
  {Winkler}}, \ and\ \bibinfo {author} {\bibfnamefont {G.}~\bibnamefont
  {Gompper}},\ }\href@noop {} {\bibfield  {journal} {\bibinfo  {journal} {Phys.
  Rev. Lett.}\ }\textbf {\bibinfo {volume} {109}},\ \bibinfo {pages} {178101}
  (\bibinfo {year} {2012})}\BibitemShut {NoStop}%
\bibitem [{\citenamefont {Jiang}\ \emph {et~al.}(2013)\citenamefont {Jiang},
  \citenamefont {Watari},\ and\ \citenamefont {Larson}}]{jian:13}%
  \BibitemOpen
  \bibfield  {author} {\bibinfo {author} {\bibfnamefont {L.}~\bibnamefont
  {Jiang}}, \bibinfo {author} {\bibfnamefont {N.}~\bibnamefont {Watari}}, \
  and\ \bibinfo {author} {\bibfnamefont {R.~G.}\ \bibnamefont {Larson}},\
  }\href {\doibase http://dx.doi.org/10.1122/1.4807857} {\bibfield  {journal}
  {\bibinfo  {journal} {J. Rheol.}\ }\textbf {\bibinfo {volume} {57}},\
  \bibinfo {pages} {1177} (\bibinfo {year} {2013})}\BibitemShut {NoStop}%
\bibitem [{\citenamefont {Noguchi}\ and\ \citenamefont
  {Gompper}(2004)}]{nogu:04}%
  \BibitemOpen
  \bibfield  {author} {\bibinfo {author} {\bibfnamefont {H.}~\bibnamefont
  {Noguchi}}\ and\ \bibinfo {author} {\bibfnamefont {G.}~\bibnamefont
  {Gompper}},\ }\href@noop {} {\bibfield  {journal} {\bibinfo  {journal} {Phys.
  Rev. Lett.}\ }\textbf {\bibinfo {volume} {93}},\ \bibinfo {pages} {258102}
  (\bibinfo {year} {2004})}\BibitemShut {NoStop}%
\bibitem [{\citenamefont {Noguchi}\ and\ \citenamefont
  {Gompper}(2005)}]{nogu:05}%
  \BibitemOpen
  \bibfield  {author} {\bibinfo {author} {\bibfnamefont {H.}~\bibnamefont
  {Noguchi}}\ and\ \bibinfo {author} {\bibfnamefont {G.}~\bibnamefont
  {Gompper}},\ }\href@noop {} {\bibfield  {journal} {\bibinfo  {journal} {Proc.
  Natl. Acad. Sci. USA}\ }\textbf {\bibinfo {volume} {102}},\ \bibinfo {pages}
  {14159} (\bibinfo {year} {2005})}\BibitemShut {NoStop}%
\bibitem [{\citenamefont {Mcwhirter}\ \emph {et~al.}(2009)\citenamefont
  {Mcwhirter}, \citenamefont {Noguchi},\ and\ \citenamefont
  {Gompper}}]{mcwh:09}%
  \BibitemOpen
  \bibfield  {author} {\bibinfo {author} {\bibfnamefont {J.~L.}\ \bibnamefont
  {Mcwhirter}}, \bibinfo {author} {\bibfnamefont {H.}~\bibnamefont {Noguchi}},
  \ and\ \bibinfo {author} {\bibfnamefont {G.}~\bibnamefont {Gompper}},\
  }\href@noop {} {\bibfield  {journal} {\bibinfo  {journal} {Proc. Natl. Acad.
  Sci. USA}\ }\textbf {\bibinfo {volume} {106}},\ \bibinfo {pages} {6039}
  (\bibinfo {year} {2009})}\BibitemShut {NoStop}%
\bibitem [{\citenamefont {Tao}\ \emph {et~al.}(2008)\citenamefont {Tao},
  \citenamefont {G{\"o}tze},\ and\ \citenamefont {Gompper}}]{tao:08}%
  \BibitemOpen
  \bibfield  {author} {\bibinfo {author} {\bibfnamefont {Y.-G.}\ \bibnamefont
  {Tao}}, \bibinfo {author} {\bibfnamefont {I.~O.}\ \bibnamefont {G{\"o}tze}},
  \ and\ \bibinfo {author} {\bibfnamefont {G.}~\bibnamefont {Gompper}},\
  }\href@noop {} {\bibfield  {journal} {\bibinfo  {journal} {J. Chem. Phys.}\
  }\textbf {\bibinfo {volume} {128}},\ \bibinfo {pages} {144902} (\bibinfo
  {year} {2008})}\BibitemShut {NoStop}%
\bibitem [{\citenamefont {Ji}\ \emph {et~al.}(2011)\citenamefont {Ji},
  \citenamefont {Jiang}, \citenamefont {Winkler},\ and\ \citenamefont
  {Gompper}}]{ji:11}%
  \BibitemOpen
  \bibfield  {author} {\bibinfo {author} {\bibfnamefont {S.}~\bibnamefont
  {Ji}}, \bibinfo {author} {\bibfnamefont {R.}~\bibnamefont {Jiang}}, \bibinfo
  {author} {\bibfnamefont {R.~G.}\ \bibnamefont {Winkler}}, \ and\ \bibinfo
  {author} {\bibfnamefont {G.}~\bibnamefont {Gompper}},\ }\href@noop {}
  {\bibfield  {journal} {\bibinfo  {journal} {J. Chem. Phys.}\ }\textbf
  {\bibinfo {volume} {135}},\ \bibinfo {pages} {134116} (\bibinfo {year}
  {2011})}\BibitemShut {NoStop}%
\bibitem [{\citenamefont {Kowalik}\ and\ \citenamefont
  {Winkler}(2013)}]{kowa:13}%
  \BibitemOpen
  \bibfield  {author} {\bibinfo {author} {\bibfnamefont {B.}~\bibnamefont
  {Kowalik}}\ and\ \bibinfo {author} {\bibfnamefont {R.~G.}\ \bibnamefont
  {Winkler}},\ }\href {\doibase 10.1063/1.4792196} {\bibfield  {journal}
  {\bibinfo  {journal} {J. Chem. Phys.}\ }\textbf {\bibinfo {volume} {138}},\
  \bibinfo {pages} {104903} (\bibinfo {year} {2013})}\BibitemShut {NoStop}%
\bibitem [{\citenamefont {R{\"u}ckner}\ and\ \citenamefont
  {Kapral}(2007)}]{ruec:07}%
  \BibitemOpen
  \bibfield  {author} {\bibinfo {author} {\bibfnamefont {G.}~\bibnamefont
  {R{\"u}ckner}}\ and\ \bibinfo {author} {\bibfnamefont {R.}~\bibnamefont
  {Kapral}},\ }\href@noop {} {\bibfield  {journal} {\bibinfo  {journal} {Phys.
  Rev. Lett.}\ }\textbf {\bibinfo {volume} {98}},\ \bibinfo {pages} {150603}
  (\bibinfo {year} {2007})}\BibitemShut {NoStop}%
\bibitem [{\citenamefont {G{\"o}tze}\ and\ \citenamefont
  {Gompper}(2010{\natexlab{b}})}]{goet:10}%
  \BibitemOpen
  \bibfield  {author} {\bibinfo {author} {\bibfnamefont {I.~O.}\ \bibnamefont
  {G{\"o}tze}}\ and\ \bibinfo {author} {\bibfnamefont {G.}~\bibnamefont
  {Gompper}},\ }\href@noop {} {\bibfield  {journal} {\bibinfo  {journal} {Phys.
  Rev. E}\ }\textbf {\bibinfo {volume} {82}},\ \bibinfo {pages} {041921}
  (\bibinfo {year} {2010}{\natexlab{b}})}\BibitemShut {NoStop}%
\bibitem [{\citenamefont {Yang}\ and\ \citenamefont {Ripoll}(2011)}]{ming:11}%
  \BibitemOpen
  \bibfield  {author} {\bibinfo {author} {\bibfnamefont {M.}~\bibnamefont
  {Yang}}\ and\ \bibinfo {author} {\bibfnamefont {M.}~\bibnamefont {Ripoll}},\
  }\href@noop {} {\bibfield  {journal} {\bibinfo  {journal} {Phys. Rev. E}\
  }\textbf {\bibinfo {volume} {84}},\ \bibinfo {pages} {061401} (\bibinfo
  {year} {2011})}\BibitemShut {NoStop}%
\bibitem [{\citenamefont {Elgeti}\ and\ \citenamefont
  {Gompper}(2009)}]{elge:09}%
  \BibitemOpen
  \bibfield  {author} {\bibinfo {author} {\bibfnamefont {J.}~\bibnamefont
  {Elgeti}}\ and\ \bibinfo {author} {\bibfnamefont {G.}~\bibnamefont
  {Gompper}},\ }\href@noop {} {\bibfield  {journal} {\bibinfo  {journal} {EPL}\
  }\textbf {\bibinfo {volume} {85}},\ \bibinfo {pages} {38002} (\bibinfo {year}
  {2009})}\BibitemShut {NoStop}%
\bibitem [{\citenamefont {Earl}\ \emph {et~al.}(2007)\citenamefont {Earl},
  \citenamefont {Pooley}, \citenamefont {Ryder}, \citenamefont {Bredberg},\
  and\ \citenamefont {Yeomans}}]{earl:07}%
  \BibitemOpen
  \bibfield  {author} {\bibinfo {author} {\bibfnamefont {D.~J.}\ \bibnamefont
  {Earl}}, \bibinfo {author} {\bibfnamefont {C.~M.}\ \bibnamefont {Pooley}},
  \bibinfo {author} {\bibfnamefont {J.~F.}\ \bibnamefont {Ryder}}, \bibinfo
  {author} {\bibfnamefont {I.}~\bibnamefont {Bredberg}}, \ and\ \bibinfo
  {author} {\bibfnamefont {J.~M.}\ \bibnamefont {Yeomans}},\ }\href@noop {}
  {\bibfield  {journal} {\bibinfo  {journal} {J. Chem. Phys.}\ }\textbf
  {\bibinfo {volume} {126}},\ \bibinfo {pages} {064703} (\bibinfo {year}
  {2007})}\BibitemShut {NoStop}%
\bibitem [{\citenamefont {Elgeti}\ \emph {et~al.}(2010)\citenamefont {Elgeti},
  \citenamefont {Kaupp},\ and\ \citenamefont {Gompper}}]{elge:10}%
  \BibitemOpen
  \bibfield  {author} {\bibinfo {author} {\bibfnamefont {J.}~\bibnamefont
  {Elgeti}}, \bibinfo {author} {\bibfnamefont {U.~B.}\ \bibnamefont {Kaupp}}, \
  and\ \bibinfo {author} {\bibfnamefont {G.}~\bibnamefont {Gompper}},\ }\href
  {\doibase 10.1016/j.bpj.2010.05.015} {\bibfield  {journal} {\bibinfo
  {journal} {Biophys. J.}\ }\textbf {\bibinfo {volume} {99}},\ \bibinfo {pages}
  {1018} (\bibinfo {year} {2010})}\BibitemShut {NoStop}%
\bibitem [{\citenamefont {Reigh}\ \emph {et~al.}(2012)\citenamefont {Reigh},
  \citenamefont {Winkler},\ and\ \citenamefont {Gompper}}]{reig:12}%
  \BibitemOpen
  \bibfield  {author} {\bibinfo {author} {\bibfnamefont {S.~Y.}\ \bibnamefont
  {Reigh}}, \bibinfo {author} {\bibfnamefont {R.~G.}\ \bibnamefont {Winkler}},
  \ and\ \bibinfo {author} {\bibfnamefont {G.}~\bibnamefont {Gompper}},\ }\href
  {\doibase 10.1039/C2SM07378A} {\bibfield  {journal} {\bibinfo  {journal}
  {Soft Matter}\ }\textbf {\bibinfo {volume} {8}},\ \bibinfo {pages} {4363}
  (\bibinfo {year} {2012})}\BibitemShut {NoStop}%
\bibitem [{\citenamefont {Theers}\ and\ \citenamefont
  {Winkler}(2013)}]{thee:13}%
  \BibitemOpen
  \bibfield  {author} {\bibinfo {author} {\bibfnamefont {M.}~\bibnamefont
  {Theers}}\ and\ \bibinfo {author} {\bibfnamefont {R.~G.}\ \bibnamefont
  {Winkler}},\ }\href {\doibase 10.1103/PhysRevE.88.023012} {\bibfield
  {journal} {\bibinfo  {journal} {Phys. Rev. E}\ }\textbf {\bibinfo {volume}
  {88}},\ \bibinfo {pages} {023012} (\bibinfo {year} {2013})}\BibitemShut
  {NoStop}%
\bibitem [{\citenamefont {Elgeti}\ \emph {et~al.}(2015)\citenamefont {Elgeti},
  \citenamefont {Winkler},\ and\ \citenamefont {Gompper}}]{elge:15}%
  \BibitemOpen
  \bibfield  {author} {\bibinfo {author} {\bibfnamefont {J.}~\bibnamefont
  {Elgeti}}, \bibinfo {author} {\bibfnamefont {R.~G.}\ \bibnamefont {Winkler}},
  \ and\ \bibinfo {author} {\bibfnamefont {G.}~\bibnamefont {Gompper}},\
  }\href@noop {} {\bibfield  {journal} {\bibinfo  {journal} {Rep. Prog. Phys.,}\
  }{\bibinfo {volume} {to appear, \textbf{arXiv:1412.2692} [physics.bio-ph]}}
  (\bibinfo {year} {2015})}\BibitemShut {NoStop}%
\bibitem [{\citenamefont {Ihle}\ \emph {et~al.}(2006)\citenamefont {Ihle},
  \citenamefont {T{\"u}zel},\ and\ \citenamefont {Kroll}}]{ihle:06}%
  \BibitemOpen
  \bibfield  {author} {\bibinfo {author} {\bibfnamefont {T.}~\bibnamefont
  {Ihle}}, \bibinfo {author} {\bibfnamefont {E.}~\bibnamefont {T{\"u}zel}}, \
  and\ \bibinfo {author} {\bibfnamefont {D.~M.}\ \bibnamefont {Kroll}},\
  }\href@noop {} {\bibfield  {journal} {\bibinfo  {journal} {Europhys. Lett.}\
  }\textbf {\bibinfo {volume} {73}},\ \bibinfo {pages} {664} (\bibinfo {year}
  {2006})}\BibitemShut {NoStop}%
\bibitem [{\citenamefont {T{\"u}zel}\ \emph {et~al.}(2007)\citenamefont
  {T{\"u}zel}, \citenamefont {Pan}, \citenamefont {Ihle},\ and\ \citenamefont
  {Kroll}}]{tuez:07}%
  \BibitemOpen
  \bibfield  {author} {\bibinfo {author} {\bibfnamefont {E.}~\bibnamefont
  {T{\"u}zel}}, \bibinfo {author} {\bibfnamefont {G.}~\bibnamefont {Pan}},
  \bibinfo {author} {\bibfnamefont {T.}~\bibnamefont {Ihle}}, \ and\ \bibinfo
  {author} {\bibfnamefont {D.~M.}\ \bibnamefont {Kroll}},\ }\href@noop {}
  {\bibfield  {journal} {\bibinfo  {journal} {EPL}\ }\textbf {\bibinfo {volume}
  {80}},\ \bibinfo {pages} {40010} (\bibinfo {year} {2007})}\BibitemShut
  {NoStop}%
\bibitem [{\citenamefont {Bird}(1994)}]{bird:94}%
  \BibitemOpen
  \bibfield  {author} {\bibinfo {author} {\bibfnamefont {G.~A.}\ \bibnamefont
  {Bird}},\ }\href@noop {} {\emph {\bibinfo {title} {Molecular Gas Dynamics and
  the Direct Simulation of Gas Flows}}}\ (\bibinfo  {publisher} {Oxford
  University Press, Oxford},\ \bibinfo {year} {1994})\BibitemShut {NoStop}%
\bibitem [{\citenamefont {Ihle}\ and\ \citenamefont
  {Kroll}(2003{\natexlab{a}})}]{ihle:03}%
  \BibitemOpen
  \bibfield  {author} {\bibinfo {author} {\bibfnamefont {T.}~\bibnamefont
  {Ihle}}\ and\ \bibinfo {author} {\bibfnamefont {D.~M.}\ \bibnamefont
  {Kroll}},\ }\href@noop {} {\bibfield  {journal} {\bibinfo  {journal} {Phys.
  Rev. E}\ }\textbf {\bibinfo {volume} {67}},\ \bibinfo {pages} {066705}
  (\bibinfo {year} {2003}{\natexlab{a}})}\BibitemShut {NoStop}%
\bibitem [{\citenamefont {Ihle}\ and\ \citenamefont {Kroll}(2001)}]{ihle:01}%
  \BibitemOpen
  \bibfield  {author} {\bibinfo {author} {\bibfnamefont {T.}~\bibnamefont
  {Ihle}}\ and\ \bibinfo {author} {\bibfnamefont {D.~M.}\ \bibnamefont
  {Kroll}},\ }\href@noop {} {\bibfield  {journal} {\bibinfo  {journal} {Phys.
  Rev. E}\ }\textbf {\bibinfo {volume} {63}},\ \bibinfo {pages} {020201(R)}
  (\bibinfo {year} {2001})}\BibitemShut {NoStop}%
\bibitem [{\citenamefont {Noguchi}\ \emph {et~al.}(2007)\citenamefont
  {Noguchi}, \citenamefont {Kikuchi},\ and\ \citenamefont {Gompper}}]{nogu:07}%
  \BibitemOpen
  \bibfield  {author} {\bibinfo {author} {\bibfnamefont {H.}~\bibnamefont
  {Noguchi}}, \bibinfo {author} {\bibfnamefont {N.}~\bibnamefont {Kikuchi}}, \
  and\ \bibinfo {author} {\bibfnamefont {G.}~\bibnamefont {Gompper}},\
  }\href@noop {} {\bibfield  {journal} {\bibinfo  {journal} {EPL}\ }\textbf
  {\bibinfo {volume} {78}},\ \bibinfo {pages} {10005} (\bibinfo {year}
  {2007})}\BibitemShut {NoStop}%
\bibitem [{\citenamefont {Huang}\ \emph
  {et~al.}(2012{\natexlab{b}})\citenamefont {Huang}, \citenamefont {Gompper},\
  and\ \citenamefont {Winkler}}]{huan:12}%
  \BibitemOpen
  \bibfield  {author} {\bibinfo {author} {\bibfnamefont {C.-C.}\ \bibnamefont
  {Huang}}, \bibinfo {author} {\bibfnamefont {G.}~\bibnamefont {Gompper}}, \
  and\ \bibinfo {author} {\bibfnamefont {R.~G.}\ \bibnamefont {Winkler}},\
  }\href {\doibase 10.1103/PhysRevE.86.056711} {\bibfield  {journal} {\bibinfo
  {journal} {Phys. Rev. E}\ }\textbf {\bibinfo {volume} {86}},\ \bibinfo
  {pages} {056711} (\bibinfo {year} {2012}{\natexlab{b}})}\BibitemShut
  {NoStop}%
\bibitem [{\citenamefont {Ihle}\ \emph {et~al.}(2005)\citenamefont {Ihle},
  \citenamefont {T{\"u}zel},\ and\ \citenamefont {Kroll}}]{ihle:05}%
  \BibitemOpen
  \bibfield  {author} {\bibinfo {author} {\bibfnamefont {T.}~\bibnamefont
  {Ihle}}, \bibinfo {author} {\bibfnamefont {E.}~\bibnamefont {T{\"u}zel}}, \
  and\ \bibinfo {author} {\bibfnamefont {D.~M.}\ \bibnamefont {Kroll}},\ }\href
  {\doibase 10.1103/PhysRevE.72.046707} {\bibfield  {journal} {\bibinfo
  {journal} {Phys. Rev. E}\ }\textbf {\bibinfo {volume} {72}},\ \bibinfo
  {pages} {046707} (\bibinfo {year} {2005})}\BibitemShut {NoStop}%
\bibitem [{\citenamefont {Kikuchi}\ \emph {et~al.}(2003)\citenamefont
  {Kikuchi}, \citenamefont {Pooley}, \citenamefont {Ryder},\ and\ \citenamefont
  {Yeomans}}]{kiku:03}%
  \BibitemOpen
  \bibfield  {author} {\bibinfo {author} {\bibfnamefont {N.}~\bibnamefont
  {Kikuchi}}, \bibinfo {author} {\bibfnamefont {C.~M.}\ \bibnamefont {Pooley}},
  \bibinfo {author} {\bibfnamefont {J.~F.}\ \bibnamefont {Ryder}}, \ and\
  \bibinfo {author} {\bibfnamefont {J.~M.}\ \bibnamefont {Yeomans}},\
  }\href@noop {} {\bibfield  {journal} {\bibinfo  {journal} {J. Chem. Phys.}\
  }\textbf {\bibinfo {volume} {119}},\ \bibinfo {pages} {6388} (\bibinfo {year}
  {2003})}\BibitemShut {NoStop}%
\bibitem [{\citenamefont {Noguchi}\ and\ \citenamefont
  {Gompper}(2008)}]{nogu:08}%
  \BibitemOpen
  \bibfield  {author} {\bibinfo {author} {\bibfnamefont {H.}~\bibnamefont
  {Noguchi}}\ and\ \bibinfo {author} {\bibfnamefont {G.}~\bibnamefont
  {Gompper}},\ }\href@noop {} {\bibfield  {journal} {\bibinfo  {journal} {Phys.
  Rev. E}\ }\textbf {\bibinfo {volume} {78}},\ \bibinfo {pages} {016706}
  (\bibinfo {year} {2008})}\BibitemShut {NoStop}%
\bibitem [{\citenamefont {Winkler}\ and\ \citenamefont
  {Huang}(2009)}]{wink:09}%
  \BibitemOpen
  \bibfield  {author} {\bibinfo {author} {\bibfnamefont {R.~G.}\ \bibnamefont
  {Winkler}}\ and\ \bibinfo {author} {\bibfnamefont {C.-C.}\ \bibnamefont
  {Huang}},\ }\href {\doibase 10.1063/1.3077860} {\bibfield  {journal}
  {\bibinfo  {journal} {J. Chem. Phys.}\ }\textbf {\bibinfo {volume} {130}},\
  \bibinfo {pages} {074907} (\bibinfo {year} {2009})}\BibitemShut {NoStop}%
\bibitem [{\citenamefont {Allen}\ and\ \citenamefont
  {Tildesley}(1987)}]{alle:87}%
  \BibitemOpen
  \bibfield  {author} {\bibinfo {author} {\bibfnamefont {M.~P.}\ \bibnamefont
  {Allen}}\ and\ \bibinfo {author} {\bibfnamefont {D.~J.}\ \bibnamefont
  {Tildesley}},\ }\href@noop {} {\emph {\bibinfo {title} {Computer Simulation
  of Liquids}}}\ (\bibinfo  {publisher} {Clarendon Press},\ \bibinfo {address}
  {Oxford},\ \bibinfo {year} {1987})\BibitemShut {NoStop}%
\bibitem [{\citenamefont {Frenkel}\ and\ \citenamefont {Smit}(2002)}]{fren:02}%
  \BibitemOpen
  \bibfield  {author} {\bibinfo {author} {\bibfnamefont {D.}~\bibnamefont
  {Frenkel}}\ and\ \bibinfo {author} {\bibfnamefont {B.}~\bibnamefont {Smit}},\
  }\href@noop {} {\emph {\bibinfo {title} {Understanding Molecular
  Simulation}}}\ (\bibinfo  {publisher} {Academic},\ \bibinfo {address} {New
  York},\ \bibinfo {year} {2002})\BibitemShut {NoStop}%
\bibitem [{\citenamefont {Andersen}(1980)}]{ande:80}%
  \BibitemOpen
  \bibfield  {author} {\bibinfo {author} {\bibfnamefont {H.~C.}\ \bibnamefont
  {Andersen}},\ }\href@noop {} {\bibfield  {journal} {\bibinfo  {journal} {J.
  Chem. Phys.}\ }\textbf {\bibinfo {volume} {72}},\ \bibinfo {pages} {2384}
  (\bibinfo {year} {1980})}\BibitemShut {NoStop}%
\bibitem [{\citenamefont {Heyes}(1983)}]{heye:83}%
  \BibitemOpen
  \bibfield  {author} {\bibinfo {author} {\bibfnamefont {D.~M.}\ \bibnamefont
  {Heyes}},\ }\href@noop {} {\bibfield  {journal} {\bibinfo  {journal} {Chem.
  Phys.}\ }\textbf {\bibinfo {volume} {82}},\ \bibinfo {pages} {285} (\bibinfo
  {year} {1983})}\BibitemShut {NoStop}%
\bibitem [{\citenamefont {Haile}\ and\ \citenamefont {Gupta}(1983)}]{hail:83}%
  \BibitemOpen
  \bibfield  {author} {\bibinfo {author} {\bibfnamefont {J.~M.}\ \bibnamefont
  {Haile}}\ and\ \bibinfo {author} {\bibfnamefont {S.}~\bibnamefont {Gupta}},\
  }\href@noop {} {\bibfield  {journal} {\bibinfo  {journal} {J. Chem. Phys.}\
  }\textbf {\bibinfo {volume} {79}},\ \bibinfo {pages} {3067} (\bibinfo {year}
  {1983})}\BibitemShut {NoStop}%
\bibitem [{\citenamefont {Berendsen}\ \emph {et~al.}(1984)\citenamefont
  {Berendsen}, \citenamefont {Postma}, \citenamefont {van Gunsteren},
  \citenamefont {DiNola},\ and\ \citenamefont {Haak}}]{bere:84}%
  \BibitemOpen
  \bibfield  {author} {\bibinfo {author} {\bibfnamefont {H.~J.~C.}\
  \bibnamefont {Berendsen}}, \bibinfo {author} {\bibfnamefont {J.~P.~M.}\
  \bibnamefont {Postma}}, \bibinfo {author} {\bibfnamefont {W.~F.}\
  \bibnamefont {van Gunsteren}}, \bibinfo {author} {\bibfnamefont
  {A.}~\bibnamefont {DiNola}}, \ and\ \bibinfo {author} {\bibfnamefont {J.~R.}\
  \bibnamefont {Haak}},\ }\href@noop {} {\bibfield  {journal} {\bibinfo
  {journal} {J. Chem. Phys.}\ }\textbf {\bibinfo {volume} {81}},\ \bibinfo
  {pages} {3684} (\bibinfo {year} {1984})}\BibitemShut {NoStop}%
\bibitem [{\citenamefont {Evans}\ and\ \citenamefont
  {Morriss}(1984)}]{evan:84}%
  \BibitemOpen
  \bibfield  {author} {\bibinfo {author} {\bibfnamefont {D.~J.}\ \bibnamefont
  {Evans}}\ and\ \bibinfo {author} {\bibfnamefont {G.~P.}\ \bibnamefont
  {Morriss}},\ }\href@noop {} {\bibfield  {journal} {\bibinfo  {journal}
  {Comput. Phys. Rep.}\ }\textbf {\bibinfo {volume} {1}},\ \bibinfo {pages}
  {297} (\bibinfo {year} {1984})}\BibitemShut {NoStop}%
\bibitem [{\citenamefont {Nos\'{e}}(1984)}]{nose:84}%
  \BibitemOpen
  \bibfield  {author} {\bibinfo {author} {\bibfnamefont {S.}~\bibnamefont
  {Nos\'{e}}},\ }\href@noop {} {\bibfield  {journal} {\bibinfo  {journal} {J.
  Chem. Phys.}\ }\textbf {\bibinfo {volume} {81}},\ \bibinfo {pages} {511}
  (\bibinfo {year} {1984})}\BibitemShut {NoStop}%
\bibitem [{\citenamefont {Hoover}(1985)}]{hoov:85}%
  \BibitemOpen
  \bibfield  {author} {\bibinfo {author} {\bibfnamefont {W.~G.}\ \bibnamefont
  {Hoover}},\ }\href@noop {} {\bibfield  {journal} {\bibinfo  {journal} {Phys.
  Rev. A}\ }\textbf {\bibinfo {volume} {31}},\ \bibinfo {pages} {1695}
  (\bibinfo {year} {1985})}\BibitemShut {NoStop}%
\bibitem [{\citenamefont {Bulgac}\ and\ \citenamefont
  {Kusnezov}(1990)}]{bulg:90}%
  \BibitemOpen
  \bibfield  {author} {\bibinfo {author} {\bibfnamefont {A.}~\bibnamefont
  {Bulgac}}\ and\ \bibinfo {author} {\bibfnamefont {D.}~\bibnamefont
  {Kusnezov}},\ }\href@noop {} {\bibfield  {journal} {\bibinfo  {journal}
  {Phys. Rev. A}\ }\textbf {\bibinfo {volume} {42}},\ \bibinfo {pages} {5045}
  (\bibinfo {year} {1990})}\BibitemShut {NoStop}%
\bibitem [{\citenamefont {Winkler}(1992)}]{wink:92.2}%
  \BibitemOpen
  \bibfield  {author} {\bibinfo {author} {\bibfnamefont {R.~G.}\ \bibnamefont
  {Winkler}},\ }\href {\doibase 10.1103/PhysRevA.45.2250} {\bibfield  {journal}
  {\bibinfo  {journal} {Phys. Rev. A}\ }\textbf {\bibinfo {volume} {45}},\
  \bibinfo {pages} {2250} (\bibinfo {year} {1992})}\BibitemShut {NoStop}%
\bibitem [{\citenamefont {Winkler}\ \emph {et~al.}(1995)\citenamefont
  {Winkler}, \citenamefont {Kraus},\ and\ \citenamefont {Reineker}}]{wink:95}%
  \BibitemOpen
  \bibfield  {author} {\bibinfo {author} {\bibfnamefont {R.~G.}\ \bibnamefont
  {Winkler}}, \bibinfo {author} {\bibfnamefont {V.}~\bibnamefont {Kraus}}, \
  and\ \bibinfo {author} {\bibfnamefont {P.}~\bibnamefont {Reineker}},\ }\href
  {\doibase 10.1063/1.468850} {\bibfield  {journal} {\bibinfo  {journal} {J.
  Chem. Phys.}\ }\textbf {\bibinfo {volume} {102}},\ \bibinfo {pages} {9018}
  (\bibinfo {year} {1995})}\BibitemShut {NoStop}%
\bibitem [{\citenamefont {Bussi}\ \emph {et~al.}(2007)\citenamefont {Bussi},
  \citenamefont {Donadio},\ and\ \citenamefont {Parrinello}}]{buss:07}%
  \BibitemOpen
  \bibfield  {author} {\bibinfo {author} {\bibfnamefont {G.}~\bibnamefont
  {Bussi}}, \bibinfo {author} {\bibfnamefont {D.}~\bibnamefont {Donadio}}, \
  and\ \bibinfo {author} {\bibfnamefont {M.}~\bibnamefont {Parrinello}},\
  }\href@noop {} {\bibfield  {journal} {\bibinfo  {journal} {J. Chem. Phys.}\
  }\textbf {\bibinfo {volume} {126}},\ \bibinfo {pages} {014101} (\bibinfo
  {year} {2007})}\BibitemShut {NoStop}%
\bibitem [{\citenamefont {Huang}\ \emph
  {et~al.}(2010{\natexlab{b}})\citenamefont {Huang}, \citenamefont {Chatterji},
  \citenamefont {Sutmann}, \citenamefont {Gompper},\ and\ \citenamefont
  {Winkler}}]{huan:10.1}%
  \BibitemOpen
  \bibfield  {author} {\bibinfo {author} {\bibfnamefont {C.-C.}\ \bibnamefont
  {Huang}}, \bibinfo {author} {\bibfnamefont {A.}~\bibnamefont {Chatterji}},
  \bibinfo {author} {\bibfnamefont {G.}~\bibnamefont {Sutmann}}, \bibinfo
  {author} {\bibfnamefont {G.}~\bibnamefont {Gompper}}, \ and\ \bibinfo
  {author} {\bibfnamefont {R.~G.}\ \bibnamefont {Winkler}},\ }\href {\doibase
  10.1016/j.jcp.2009.09.024} {\bibfield  {journal} {\bibinfo  {journal} {J.
  Comput. Phys.}\ }\textbf {\bibinfo {volume} {229}},\ \bibinfo {pages} {168}
  (\bibinfo {year} {2010}{\natexlab{b}})}\BibitemShut {NoStop}%
\bibitem [{\citenamefont {Schwarzer}(1995)}]{schw:95}%
  \BibitemOpen
  \bibfield  {author} {\bibinfo {author} {\bibfnamefont {S.}~\bibnamefont
  {Schwarzer}},\ }\href@noop {} {\bibfield  {journal} {\bibinfo  {journal}
  {Phys. Rev. E}\ }\textbf {\bibinfo {volume} {52}},\ \bibinfo {pages} {6461}
  (\bibinfo {year} {1995})}\BibitemShut {NoStop}%
\bibitem [{\citenamefont {Evans}\ and\ \citenamefont
  {Morriss}(1986)}]{evan:86}%
  \BibitemOpen
  \bibfield  {author} {\bibinfo {author} {\bibfnamefont {D.~J.}\ \bibnamefont
  {Evans}}\ and\ \bibinfo {author} {\bibfnamefont {G.~P.}\ \bibnamefont
  {Morriss}},\ }\href {\doibase 10.1103/PhysRevLett.56.2172} {\bibfield
  {journal} {\bibinfo  {journal} {Phys. Rev. Lett.}\ }\textbf {\bibinfo
  {volume} {56}},\ \bibinfo {pages} {2172} (\bibinfo {year}
  {1986})}\BibitemShut {NoStop}%
\bibitem [{\citenamefont {Bolintineanu}\ \emph {et~al.}(2012)\citenamefont
  {Bolintineanu}, \citenamefont {Lechman}, \citenamefont {Plimpton},\ and\
  \citenamefont {Grest}}]{boli:12}%
  \BibitemOpen
  \bibfield  {author} {\bibinfo {author} {\bibfnamefont {D.~S.}\ \bibnamefont
  {Bolintineanu}}, \bibinfo {author} {\bibfnamefont {J.~B.}\ \bibnamefont
  {Lechman}}, \bibinfo {author} {\bibfnamefont {S.~J.}\ \bibnamefont
  {Plimpton}}, \ and\ \bibinfo {author} {\bibfnamefont {G.~S.}\ \bibnamefont
  {Grest}},\ }\href {\doibase 10.1103/PhysRevE.86.066703} {\bibfield  {journal}
  {\bibinfo  {journal} {Phys. Rev. E}\ }\textbf {\bibinfo {volume} {86}},\
  \bibinfo {pages} {066703} (\bibinfo {year} {2012})}\BibitemShut {NoStop}%
\bibitem [{\citenamefont {Boon}\ and\ \citenamefont {Yip}(1980)}]{boon:80}%
  \BibitemOpen
  \bibfield  {author} {\bibinfo {author} {\bibfnamefont {J.~P.}\ \bibnamefont
  {Boon}}\ and\ \bibinfo {author} {\bibfnamefont {S.}~\bibnamefont {Yip}},\
  }\href@noop {} {\emph {\bibinfo {title} {Molecular Hydrodynamics}}}\
  (\bibinfo  {publisher} {Dover},\ \bibinfo {address} {New York},\ \bibinfo
  {year} {1980})\BibitemShut {NoStop}%
\bibitem [{\citenamefont {Hansen}\ and\ \citenamefont
  {McDonald}(1986)}]{hans:86}%
  \BibitemOpen
  \bibfield  {author} {\bibinfo {author} {\bibfnamefont {J.-P.}\ \bibnamefont
  {Hansen}}\ and\ \bibinfo {author} {\bibfnamefont {I.~R.}\ \bibnamefont
  {McDonald}},\ }\href@noop {} {\emph {\bibinfo {title} {Theory of Simple
  Liquids}}}\ (\bibinfo  {publisher} {Academic Press},\ \bibinfo {address}
  {London},\ \bibinfo {year} {1986})\BibitemShut {NoStop}%
\bibitem [{\citenamefont {Inoue}\ \emph {et~al.}(2002)\citenamefont {Inoue},
  \citenamefont {Chen},\ and\ \citenamefont {Ohashi}}]{inou:02}%
  \BibitemOpen
  \bibfield  {author} {\bibinfo {author} {\bibfnamefont {Y.}~\bibnamefont
  {Inoue}}, \bibinfo {author} {\bibfnamefont {Y.}~\bibnamefont {Chen}}, \ and\
  \bibinfo {author} {\bibfnamefont {H.}~\bibnamefont {Ohashi}},\ }\href@noop {}
  {\bibfield  {journal} {\bibinfo  {journal} {J. Stat. Phys.}\ }\textbf
  {\bibinfo {volume} {107}},\ \bibinfo {pages} {85} (\bibinfo {year}
  {2002})}\BibitemShut {NoStop}%
\bibitem [{\citenamefont {Lees}\ and\ \citenamefont {Edwards}(1972)}]{lees:72}%
  \BibitemOpen
  \bibfield  {author} {\bibinfo {author} {\bibfnamefont {A.~W.}\ \bibnamefont
  {Lees}}\ and\ \bibinfo {author} {\bibfnamefont {S.~F.}\ \bibnamefont
  {Edwards}},\ }\href@noop {} {\bibfield  {journal} {\bibinfo  {journal} {J.
  Phys. C}\ }\textbf {\bibinfo {volume} {5}},\ \bibinfo {pages} {1921}
  (\bibinfo {year} {1972})}\BibitemShut {NoStop}%
\bibitem [{\citenamefont {Pooley}\ and\ \citenamefont
  {Yeomans}(2005)}]{pool:05}%
  \BibitemOpen
  \bibfield  {author} {\bibinfo {author} {\bibfnamefont {C.~M.}\ \bibnamefont
  {Pooley}}\ and\ \bibinfo {author} {\bibfnamefont {J.~M.}\ \bibnamefont
  {Yeomans}},\ }\href@noop {} {\bibfield  {journal} {\bibinfo  {journal} {J.
  Phys. Chem. B}\ }\textbf {\bibinfo {volume} {109}},\ \bibinfo {pages} {6505}
  (\bibinfo {year} {2005})}\BibitemShut {NoStop}%
\bibitem [{\citenamefont {Ripoll}\ \emph {et~al.}(2005)\citenamefont {Ripoll},
  \citenamefont {Mussawisade}, \citenamefont {Winkler},\ and\ \citenamefont
  {Gompper}}]{ripo:05}%
  \BibitemOpen
  \bibfield  {author} {\bibinfo {author} {\bibfnamefont {M.}~\bibnamefont
  {Ripoll}}, \bibinfo {author} {\bibfnamefont {K.}~\bibnamefont {Mussawisade}},
  \bibinfo {author} {\bibfnamefont {R.~G.}\ \bibnamefont {Winkler}}, \ and\
  \bibinfo {author} {\bibfnamefont {G.}~\bibnamefont {Gompper}},\ }\href@noop
  {} {\bibfield  {journal} {\bibinfo  {journal} {Phys. Rev. E}\ }\textbf
  {\bibinfo {volume} {72}},\ \bibinfo {pages} {016701} (\bibinfo {year}
  {2005})}\BibitemShut {NoStop}%
\bibitem [{\citenamefont {T{\"u}zel}\ \emph {et~al.}(2006)\citenamefont
  {T{\"u}zel}, \citenamefont {Ihle},\ and\ \citenamefont {Kroll}}]{tuez:06}%
  \BibitemOpen
  \bibfield  {author} {\bibinfo {author} {\bibfnamefont {E.}~\bibnamefont
  {T{\"u}zel}}, \bibinfo {author} {\bibfnamefont {T.}~\bibnamefont {Ihle}}, \
  and\ \bibinfo {author} {\bibfnamefont {D.~M.}\ \bibnamefont {Kroll}},\
  }\href@noop {} {\bibfield  {journal} {\bibinfo  {journal} {Phys. Rev. E}\
  }\textbf {\bibinfo {volume} {74}},\ \bibinfo {pages} {056702} (\bibinfo
  {year} {2006})}\BibitemShut {NoStop}%
\bibitem [{\citenamefont {Westphal}\ \emph {et~al.}(2014)\citenamefont
  {Westphal}, \citenamefont {Singh}, \citenamefont {Huang}, \citenamefont
  {Gompper},\ and\ \citenamefont {Winkler}}]{west:14}%
  \BibitemOpen
  \bibfield  {author} {\bibinfo {author} {\bibfnamefont {E.}~\bibnamefont
  {Westphal}}, \bibinfo {author} {\bibfnamefont {S.~P.}\ \bibnamefont {Singh}},
  \bibinfo {author} {\bibfnamefont {C.-C.}\ \bibnamefont {Huang}}, \bibinfo
  {author} {\bibfnamefont {G.}~\bibnamefont {Gompper}}, \ and\ \bibinfo
  {author} {\bibfnamefont {R.~G.}\ \bibnamefont {Winkler}},\ }\href {\doibase
  10.1016/j.cpc.2013.10.004} {\bibfield  {journal} {\bibinfo  {journal}
  {Comput. Phys. Comm.}\ }\textbf {\bibinfo {volume} {185}},\ \bibinfo {pages}
  {495} (\bibinfo {year} {2014})}\BibitemShut {NoStop}%
\bibitem [{\citenamefont {Hecht}\ \emph {et~al.}(2006)\citenamefont {Hecht},
  \citenamefont {Harting}, \citenamefont {Bier}, \citenamefont {Reinshagen},\
  and\ \citenamefont {Herrmann}}]{hech:06}%
  \BibitemOpen
  \bibfield  {author} {\bibinfo {author} {\bibfnamefont {M.}~\bibnamefont
  {Hecht}}, \bibinfo {author} {\bibfnamefont {J.}~\bibnamefont {Harting}},
  \bibinfo {author} {\bibfnamefont {M.}~\bibnamefont {Bier}}, \bibinfo {author}
  {\bibfnamefont {J.}~\bibnamefont {Reinshagen}}, \ and\ \bibinfo {author}
  {\bibfnamefont {H.~J.}\ \bibnamefont {Herrmann}},\ }\href {\doibase
  10.1103/PhysRevE.74.021403} {\bibfield  {journal} {\bibinfo  {journal} {Phys.
  Rev. E}\ }\textbf {\bibinfo {volume} {74}},\ \bibinfo {pages} {021403}
  (\bibinfo {year} {2006})}\BibitemShut {NoStop}%
\bibitem [{\citenamefont {Hecht}\ \emph
  {et~al.}(2007{\natexlab{a}})\citenamefont {Hecht}, \citenamefont {Harting},\
  and\ \citenamefont {Herrmann}}]{hech:07}%
  \BibitemOpen
  \bibfield  {author} {\bibinfo {author} {\bibfnamefont {M.}~\bibnamefont
  {Hecht}}, \bibinfo {author} {\bibfnamefont {J.}~\bibnamefont {Harting}}, \
  and\ \bibinfo {author} {\bibfnamefont {H.~J.}\ \bibnamefont {Herrmann}},\
  }\href {\doibase 10.1103/PhysRevE.75.051404} {\bibfield  {journal} {\bibinfo
  {journal} {Phys. Rev. E}\ }\textbf {\bibinfo {volume} {75}},\ \bibinfo
  {pages} {051404} (\bibinfo {year} {2007}{\natexlab{a}})}\BibitemShut
  {NoStop}%
\bibitem [{\citenamefont {Hecht}\ \emph
  {et~al.}(2007{\natexlab{b}})\citenamefont {Hecht}, \citenamefont {Harting},\
  and\ \citenamefont {Herrmann}}]{hech:07.1}%
  \BibitemOpen
  \bibfield  {author} {\bibinfo {author} {\bibfnamefont {M.}~\bibnamefont
  {Hecht}}, \bibinfo {author} {\bibfnamefont {J.}~\bibnamefont {Harting}}, \
  and\ \bibinfo {author} {\bibfnamefont {H.~J.}\ \bibnamefont {Herrmann}},\
  }\href@noop {} {\bibfield  {journal} {\bibinfo  {journal} {Int. J. Mod. Phys.
  C}\ }\textbf {\bibinfo {volume} {18}},\ \bibinfo {pages} {501} (\bibinfo
  {year} {2007}{\natexlab{b}})}\BibitemShut {NoStop}%
\bibitem [{\citenamefont {Ihle}\ and\ \citenamefont
  {Kroll}(2003{\natexlab{b}})}]{ihle:03.1}%
  \BibitemOpen
  \bibfield  {author} {\bibinfo {author} {\bibfnamefont {T.}~\bibnamefont
  {Ihle}}\ and\ \bibinfo {author} {\bibfnamefont {D.~M.}\ \bibnamefont
  {Kroll}},\ }\href@noop {} {\bibfield  {journal} {\bibinfo  {journal} {Phys.
  Rev. E}\ }\textbf {\bibinfo {volume} {67}},\ \bibinfo {pages} {066706}
  (\bibinfo {year} {2003}{\natexlab{b}})}\BibitemShut {NoStop}%
\bibitem [{\citenamefont {D{\"u}nweg}(1993)}]{duen:93.1}%
  \BibitemOpen
  \bibfield  {author} {\bibinfo {author} {\bibfnamefont {B.}~\bibnamefont
  {D{\"u}nweg}},\ }\href {\doibase http://dx.doi.org/10.1063/1.465444}
  {\bibfield  {journal} {\bibinfo  {journal} {J. Chem. Phys.}\ }\textbf
  {\bibinfo {volume} {99}},\ \bibinfo {pages} {6977} (\bibinfo {year}
  {1993})}\BibitemShut {NoStop}%
\bibitem [{\citenamefont {Landau}\ and\ \citenamefont
  {Lifshitz}(1959)}]{land:59}%
  \BibitemOpen
  \bibfield  {author} {\bibinfo {author} {\bibfnamefont {L.~D.}\ \bibnamefont
  {Landau}}\ and\ \bibinfo {author} {\bibfnamefont {E.~M.}\ \bibnamefont
  {Lifshitz}},\ }\href@noop {} {\emph {\bibinfo {title} {Fluid Mechanics}}}\
  (\bibinfo  {publisher} {Pergamon Press},\ \bibinfo {address} {London},\
  \bibinfo {year} {1959})\BibitemShut {NoStop}%
\bibitem [{\citenamefont {Theers}\ and\ \citenamefont
  {Winkler}(2014)}]{thee:14}%
  \BibitemOpen
  \bibfield  {author} {\bibinfo {author} {\bibfnamefont {M.}~\bibnamefont
  {Theers}}\ and\ \bibinfo {author} {\bibfnamefont {R.~G.}\ \bibnamefont
  {Winkler}},\ }\href {\doibase 10.1039/C4SM00770K} {\bibfield  {journal}
  {\bibinfo  {journal} {Soft Matter}\ }\textbf {\bibinfo {volume} {10}},\
  \bibinfo {pages} {5894} (\bibinfo {year} {2014})}\BibitemShut {NoStop}%
\bibitem [{\citenamefont {Alder}\ and\ \citenamefont
  {Wainwright}(1970)}]{alde:70}%
  \BibitemOpen
  \bibfield  {author} {\bibinfo {author} {\bibfnamefont {B.~J.}\ \bibnamefont
  {Alder}}\ and\ \bibinfo {author} {\bibfnamefont {T.~E.}\ \bibnamefont
  {Wainwright}},\ }\href {\doibase 10.1103/PhysRevA.1.18} {\bibfield  {journal}
  {\bibinfo  {journal} {Phys. Rev. A}\ }\textbf {\bibinfo {volume} {1}},\
  \bibinfo {pages} {18} (\bibinfo {year} {1970})}\BibitemShut {NoStop}%
\bibitem [{\citenamefont {Zwanzig}\ and\ \citenamefont
  {Bixon}(1970)}]{zwan:70}%
  \BibitemOpen
  \bibfield  {author} {\bibinfo {author} {\bibfnamefont {R.}~\bibnamefont
  {Zwanzig}}\ and\ \bibinfo {author} {\bibfnamefont {M.}~\bibnamefont
  {Bixon}},\ }\href@noop {} {\bibfield  {journal} {\bibinfo  {journal} {Phys.
  Rev. A}\ }\textbf {\bibinfo {volume} {2}},\ \bibinfo {pages} {2005} (\bibinfo
  {year} {1970})}\BibitemShut {NoStop}%
\bibitem [{\citenamefont {H\'{i}jar}\ and\ \citenamefont
  {Sutmann}(2011)}]{hija:11}%
  \BibitemOpen
  \bibfield  {author} {\bibinfo {author} {\bibfnamefont {H.}~\bibnamefont
  {H\'{i}jar}}\ and\ \bibinfo {author} {\bibfnamefont {G.}~\bibnamefont
  {Sutmann}},\ }\href@noop {} {\bibfield  {journal} {\bibinfo  {journal} {Phys.
  Rev. E}\ }\textbf {\bibinfo {volume} {83}},\ \bibinfo {pages} {046708}
  (\bibinfo {year} {2011})}\BibitemShut {NoStop}%
\bibitem [{\citenamefont {Poblete}\ \emph {et~al.}(2014)\citenamefont
  {Poblete}, \citenamefont {Wysocki}, \citenamefont {Gompper},\ and\
  \citenamefont {Winkler}}]{pobl:14}%
  \BibitemOpen
  \bibfield  {author} {\bibinfo {author} {\bibfnamefont {S.}~\bibnamefont
  {Poblete}}, \bibinfo {author} {\bibfnamefont {A.}~\bibnamefont {Wysocki}},
  \bibinfo {author} {\bibfnamefont {G.}~\bibnamefont {Gompper}}, \ and\
  \bibinfo {author} {\bibfnamefont {R.~G.}\ \bibnamefont {Winkler}},\ }\href
  {\doibase 10.1103/PhysRevE.90.033314} {\bibfield  {journal} {\bibinfo
  {journal} {Physical Review E}\ }\textbf {\bibinfo {volume} {90}},\ \bibinfo
  {pages} {033314} (\bibinfo {year} {2014})}\BibitemShut {NoStop}%
\end{thebibliography}

%

\end{document}